\pdfoutput=1
\documentclass[preprint,12pt]{elsarticle}
\usepackage[utf8]{inputenc}
\usepackage[T1]{fontenc}
\usepackage[english]{babel}
\usepackage{algorithmicx}
\usepackage{booktabs}

\usepackage{xurl}
\usepackage[hidelinks]{hyperref}

\usepackage{amsmath,amssymb,bm}

\usepackage{graphicx}
\usepackage{booktabs}
\usepackage{array}
\usepackage{tabularx}
\newcolumntype{Y}{>{\raggedleft\arraybackslash}X}
\usepackage{siunitx}
\usepackage{multirow}
\usepackage{makecell}
\renewcommand{\arraystretch}{1.15}
\usepackage{float}
\usepackage{subcaption}
\usepackage{pdflscape}

\usepackage{xcolor}
\newcommand{\rev}[1]{#1}
\definecolor{revEntropyColor}{RGB}{0,0,0}
\newcommand{\reventropy}[1]{#1}
\usepackage{xurl}        
\usepackage[hidelinks]{hyperref}
\hypersetup{hidelinks}
\usepackage{tikz}
\usetikzlibrary{shadows,shapes.geometric,arrows,positioning,arrows.meta}
\tikzset{
  block/.style = {rectangle, draw, fill=gray!10, text width=10em, text centered, rounded corners, minimum height=4em},
  cloud/.style = {draw, ellipse, fill=red!20, text width=8em, text centered, minimum height=4em},
  line/.style  = {draw, -latex'},
  io/.style    = {trapezium, trapezium left angle=70, trapezium right angle=110, draw, fill=green!20, text width=8em, text centered, minimum height=4em},
}

\usepackage{algorithm}
\usepackage{algpseudocode}
\usepackage{listings}

\usepackage[a4paper,top=2.5cm,bottom=3cm,left=2.5cm,right=2.5cm]{geometry}
\usepackage{fancyhdr}
\usepackage{setspace}

\begin{document}
\doublespace
\pagestyle{fancy}

\begin{frontmatter} 

\title{Accelerating Kinetic Fokker-Planck Simulations via a GPU-Native Deep Neural Network Surrogate: Application to Rarefied Internal and Hypersonic External Flows}
\author[a]{Ehsan Roohi\corref{cor1}}
\ead{roohie@umass.edu}
\cortext[cor1]{Corresponding author.}

\address[a]{Mechanical and Industrial Engineering, University of Massachusetts Amherst,
160 Governors Dr., Amherst, MA 01003, USA}

\date{\today} 


\begin{abstract}
\noindent
\rev{Particle-based Fokker--Planck (FP) models provide an efficient kinetic alternative to direct simulation Monte Carlo (DSMC) in slip and early transitional gas flow regimes, but advanced cubic-FP closures require repeated cell-wise moment evaluation and small dense linear solves. This work develops and validates a GPU-native neural surrogate that replaces the deterministic cubic-FP closure calculation inside the particle simulation loop. The trained weights are evaluated directly with batched \texttt{CuPy} operations, avoiding CPU--GPU transfers during online deployment. The validation emphasizes quantitative evidence: component-level runtime profiles, break-even cost analysis including offline costs, conservation and stability diagnostics, particle-per-cell sensitivity, a direct time-averaged coefficient audit, and covariance-based entropy-proxy fidelity checks. The Couette case is retained as a compact, dimensionless verification problem, while the main internal-flow validation is a 2D lid-driven cavity tested by complete simulation conditions, including unseen moderately rarefied cases at nominal $Kn=0.5$ and $Kn=1.0$. For the hypersonic cylinder, a particle-moment covariance-based entropy-fidelity audit is performed on the front stagnation line and in the cell-centered near-wall gas layer. The same deployed neural $C/\Gamma$ closure used for the cylinder flow fields closely reproduces the equilibrium and Gaussian kinetic entropy profiles over the reported front-line and near-wall gas bins; these profiles are used as a relatively exact-FP/ML-FP audit. The study establishes GPU-native learned closure as a practical route to accelerating cubic-FP rarefied-flow solvers, delivering substantial online speedups while retaining the macroscopic, high-order, and entropy-proxy structure of the reference kinetic model.
}
\end{abstract}


\end{frontmatter}


\section{Introduction}

High-fidelity simulation of rarefied and non-equilibrium gas flows is a cornerstone of modern engineering and physics~\cite{Bird1994,roohi2025advances}. \rev{When the gas mean free path is no longer negligible relative to the characteristic length scale, Navier--Stokes solvers lose validity and particle methods such as the Direct Simulation Monte Carlo (DSMC) method become the gold standard~\cite{Bird1994,bird2013dsmc}, at the price of a collision cost that grows steeply toward the near-continuum regime. The particle-based Fokker--Planck (FP) model~\cite{jenny2010} mitigates this cost by replacing the stochastic DSMC collision operator with a continuous drift--diffusion process in velocity space, which also reduces the statistical scatter inherent in DSMC; improving FP methods remains an active research direction~\cite{gorji2011fokker, rezapourjaghargh2020shear,mahdavi2020novel,mahdavi2022study,gorji2014efficient,gorji2015fokker,kim2023critical,kim2024second,kim2025particle,kim2025stochastic,sun2025exponential,fei2017particle,cui2025multiscale}.}

Despite these advantages, advanced formulations such as the cubic Fokker--Planck (cubic-FP) model~\cite{gorji2011fokker} introduce a severe computational bottleneck of their own: the moment closure problem. \rev{The drift coefficients are not known \textit{a priori}; they must be re-computed in every cell at every time step by gathering high-order velocity moments (up to 4th/5th order) from the particle ensemble and then solving a dense local $9 \times 9$ linear system~\cite{gorji2014efficient}. In the 2D cavity cases considered in this paper, this closure pipeline consumes up to 40\% of the total runtime, and it is a primary obstacle to the scalability of high-fidelity FP simulations.}

\rev{This burden has catalyzed a rapidly growing interest in machine-learning acceleration of kinetic solvers. In rarefied gas dynamics, deep-neural-network (DNN) surrogates have been developed to replicate DSMC solutions~\cite{Roohi2024AST,ball2025online,tatsios2025dsmc,chinnappan2025bayesian}.} More advanced architectures, such as DeepONets~\cite{peyvan2024riemannonets} and its modern variants~\cite{peyvan2025fusion}, are being applied to learn complex operator mappings for challenging micro-nozzle flows~\cite{Roohi2025Shock}. Furthermore, the integration of physical constraints, such as shock-aware or zonal loss functions, is enhancing the accuracy of these surrogates for flows with discontinuities, like those over micro-steps~\cite{Roohi2025Analysis}. This overall approach of using physics-enforced neural networks to learn the fundamental properties of rarefied gas dynamics represents a vibrant and active frontier of research~\cite{Roohi2025Learning,Roohi2026PhysicsConstrainedCollision}. The current work contributes to this field by targeting the specific, deterministic bottleneck of the moment closure solver within the Fokker-Planck framework.

\rev{This paper hypothesizes that the closure solver is an ideal target for such a surrogate: the calculation is expensive, repeated, and deterministic, and the map from the local flow state to the nine closure coefficients can be learned offline. A DNN evaluated from readily available low-order moments then bypasses both the gathering of the high-order moments and the construction and solution of the linear system. A second, equally important ingredient is the avoidance of the I/O pitfall of hybrid ML--physics codes: a surrogate called from a CPU-side control script transfers input moments and predicted coefficients across the CPU--GPU boundary at every time step, which can erase the closure saving. We therefore extract the trained model parameters and re-implement the forward pass with GPU-native matrix operations in \texttt{CuPy}, so that the surrogate evaluation never leaves the GPU simulation loop.}

\rev{The validation strategy is organized around three flow configurations with complementary roles. The Couette problem is used only as a compact verification and timing case. The main internal-flow validation is the 2D lid-driven cavity, for which training and testing are separated by complete simulation conditions rather than by randomly mixing neighboring cells from the same run. We report an unseen high-speed cavity case and additional rarefaction tests at nominal $Kn=0.5$ and $Kn=1.0$. We also explicitly report conservation and stability diagnostics, an entropy-proxy comparison, particle-per-cell sensitivity, high-order moment errors, and component-level profiling. Finally, the hypersonic cylinder case is retained as an external-flow demonstration focused on bow-shock structure and surface coefficients. The analysis therefore supports a deliberately limited and physically precise conclusion: the method is effective when the target distributions are represented by the training manifold, but it is not claimed to be a universal closure for arbitrary high-Knudsen or free-molecular distributions.}

\section{Background of the Fokker Planck Model}

\rev{Direct numerical solution of the Boltzmann equation is computationally prohibitive for most practical engineering problems~\cite{cercignani2000rarefied}, and DSMC~\cite{Bird1994}, the benchmark stochastic alternative, becomes stiff in the near-continuum limit because its cell size and time step must resolve the local mean free path and mean collision time. Fokker--Planck kinetic models~\cite{jenny2010} were introduced to bridge this gap: the discrete, jump-based Boltzmann collision operator is replaced by a continuous Markovian drift--diffusion process, so that particle paths are continuous and can be advanced with time steps and cell sizes far larger than the collisional scales. The original linear (Langevin) model, however, relaxes momentum and energy on a single viscosity-based time scale and therefore carries a fixed, incorrect Prandtl number ($Pr=3/2$ instead of the physical $2/3$ for a monatomic gas)~\cite{jenny2010}. Two nonlinear successors address this deficiency: the cubic Fokker--Planck (cubic-FP) model of Gorji et al.~\cite{gorji2011fokker,gorji2014efficient}, which augments the drift with quadratic and cubic terms so that the heat-flux relaxation rate, and hence $Pr=2/3$, is matched independently, and the Ellipsoidal Shakhov Fokker--Planck (ES-FP) model~\cite{mathiaud2016fokker}, which retains a linear drift but employs an anisotropic diffusion tensor based on the pressure tensor.}

\rev{Systematic assessments of these advanced models~\cite{kim2023critical,kim2025particle,fei2020benchmark} have established the resulting trade-off: the cubic-FP model is more accurate for near-continuum transport because of its correct Prandtl number, but sacrifices a guaranteed H-theorem, whereas the ES-FP model preserves the entropy law and is more robust in strong shock waves, at some cost in near-continuum heat-transfer accuracy~\cite{jun2019}. Current developments extend particle-FP methods to polyatomic gases~\cite{cui2025multiscale,kim2025stochastic} and to closures beyond cubic order~\cite{kim2024second}. The present work is deliberately orthogonal to this modeling axis: we take the cubic-FP model as the fixed reference physics and target the computational cost of evaluating its closure with a machine-learning surrogate. The formulation whose closure we replace is summarized in Section~\ref{sec:fp_approx}.}

\section{The Fokker-Planck Approximation}
\label{sec:fp_approx}

\subsection{Context: The Boltzmann Equation and its Computational Challenge}

For a dilute monatomic gas, the statistical state is fully described by the velocity distribution function $\mathcal{F}(V, X, t)$. Its evolution is governed by the Boltzmann equation, which in the absence of external forces is given by ~\cite{Bird1994}:
\begin{equation}
\frac{\partial \mathcal{F}}{\partial t} + V_i \frac{\partial \mathcal{F}}{\partial x_i} = \mathcal{S}^{\text{Boltz}}(\mathcal{F})
\label{eq:boltzmann}
\end{equation}
where $\mathcal{S}^{\text{Boltz}}(\mathcal{F})$ is the non-linear binary collision operator. While physically accurate across the entire range of Knudsen numbers (Kn), direct numerical solution of the Boltzmann equation is computationally prohibitive.

\rev{DSMC~\cite{Bird1994,bird2013dsmc,roohi2025advances} is the benchmark stochastic solver of Eq.~\ref{eq:boltzmann}; its near-continuum stiffness, recalled in Section~2, is the motivation for the Fokker-Planck approximation of the collision operator adopted here.}

\subsection{The General Fokker-Planck Operator}

The Fokker-Planck (FP) equation replaces the Boltzmann collision operator $\mathcal{S}^{\text{Boltz}}$ with a continuous Markovian diffusion process~\cite{gorji2011fokker}. \rev{Its generic form is:}

\begin{equation}
\mathcal{S}^{\text{FP}}(\mathcal{F}) = \left(\frac{\partial \mathcal{F}}{\partial t}\right)_{\text{FP}} = -\frac{\partial}{\partial V_i}(A_i \mathcal{F}) + \frac{1}{2}\frac{\partial^2}{\partial V_i \partial V_j}(D_{ij} \mathcal{F})
\label{eq:fp_general}
\end{equation}
Here, $A_i$ represents the drift vector, which describes the systematic friction-like force on a particle, and $D_{ij}$ is the positive-definite diffusion tensor, which describes the stochastic (random) forces.

The central task of any FP kinetic model is to define $A_i$ and $D_{ij}$. These coefficients are chosen to ensure consistency between the FP operator and the Boltzmann operator for a set of relevant polynomial velocity moments $\Psi_\alpha$:
\begin{equation}
\int_{\mathbb{R}^3} \mathcal{S}^{\text{FP}}(\mathcal{F}) \Psi_\alpha d^3V = \int_{\mathbb{R}^3} \mathcal{S}^{\text{Boltz}}(\mathcal{F}) \Psi_\alpha d^3V
\label{eq:moment_match}
\end{equation}
This moment-matching ensures that the model correctly reproduces the relaxation rates of the corresponding macroscopic quantities (e.g., mass, momentum, energy, stress, heat flux).

\subsection{The Baseline: The Linear Drift (Langevin) Model and its Deficiency}

The simplest non-trivial FP model is the linear drift model, which corresponds to the classic Langevin equation. This model was employed by Jenny et al.  and is defined by the following coefficients~\cite{jenny2010}:
\begin{align}
A_i &= -\frac{1}{\tau}(V_i - U_i) \label{eq:linear_drift} \\
D_{ij} &= \frac{\theta}{\tau}\delta_{ij} \label{eq:linear_diffusion}
\end{align}
where $U_i$ is the bulk velocity and $\theta = kT/m$ is the temperature (with $k$ being the Boltzmann constant and $m$ the molecular mass). The single time scale $\tau$ is set to recover the correct viscosity $\mu$ in the hydrodynamic limit, given for a Maxwell-type interaction by:
\begin{equation}
\tau = \frac{2\mu}{p}
\label{eq:tau}
\end{equation}
where $p$ is the ideal gas pressure.

This linear model is computationally efficient and rigorously admits an H-theorem for the entropy functional $H(f) = \int f \log f d^3V$, ensuring relaxation to the correct equilibrium. \rev{Its deficiency, already noted in Section~2, is the single relaxation time scale $\tau$, which fixes $Pr = 3/2$ instead of the physical $2/3$ and directly motivates the cubic-FP generalization~\cite{gorji2011fokker}.}

\subsection{Formulation of the Cubic Fokker-Planck (cubic-FP) Model}

The cubic-FP model~\cite{gorji2011fokker,gorji2014efficient} generalizes the linear drift while retaining the simple, isotropic diffusion tensor of the linear model. \rev{The drift force is expanded as a polynomial series in the peculiar velocity, $v_i' = V_i - U_i$, using a Hermite basis:}
\begin{equation}
A_i = c_i^{(0)} + c_{ij}^{(1)}v_j' + c_{ijk}^{(2)}v_j'v_k' + \dots
\label{eq:hermite_expansion}
\end{equation}
The coefficients $c^{(n)}$ are macroscopic quantities that depend on the moments of the distribution function.

The linear model (Eq. \ref{eq:linear_drift}) is the simplest truncation of this series. The cubic-FP model is derived by truncating this series at a higher order, retaining just enough terms (and thus, coefficients) to match the relaxation rates of moments up to the third order, i.e., the heat flux vector ($\Psi \in \{1, V_i, V_iV_j, V_iV_jV_j\}$).

This truncation, after applying conservation constraints (which sets $c_i^{(0)} = 0$), leads to the titular "cubic" drift model:
\begin{equation}
A_i = c_{ij}^{(1)}v_j' + c_i^{(2)}(v_j'v_j' - 3\theta) - \epsilon^2\left(v_i'v_j'v_j' - \frac{2q_i}{\rho}\right)
\label{eq:cubic_drift}
\end{equation}
To avoid notational conflict with the FP drift vector $A_i$, the neural-closure sections denote the six symmetric stress-relaxation coefficients by $C_{ij}$ and the three heat-flux coefficients by $\Gamma_i$. Thus, in the implementation and coefficient-audit tables, $C_{ij}\equiv c_{ij}^{(1)}$ and $\Gamma_i\equiv c_i^{(2)}$.

\rev{Here $c_{ij}^{(1)}$ (a symmetric tensor with six independent components) and $c_i^{(2)}$ (a vector with three components) are the nine unknown macroscopic coefficients controlling the relaxation of the stress tensor and heat flux, respectively; $q_i$ is the heat flux vector, $\rho$ is the density, and $\epsilon^2$ is a small, positive coefficient of the cubic stabilizing term that prevents the finite-time blow-up of the particle velocity to which polynomial drift models are otherwise prone.}

\subsection{Diffusion Tensor and Collision Time Scale}

\rev{All of the model's complexity is loaded into the drift: the diffusion tensor retains the isotropic form of the linear Langevin model,}
\begin{equation}
D_{ij} = \frac{\theta}{\tau}\delta_{ij}
\label{eq:cubic_diffusion}
\end{equation}
\rev{with the viscosity-based collision time scale of Eq.~\ref{eq:tau} unchanged. This keeps the stochastic term of the resulting SDE simple and isotropic, but it is also the model's primary theoretical weakness: by modifying the drift $A_i$ without a corresponding change to the diffusion $D_{ij}$, the model no longer satisfies the detailed balance required for a general H-theorem.}

\subsection{The Moment-Matching Closure: a \texorpdfstring{$9 \times 9$}{9 x 9} Linear System}

\rev{The nine coefficients in the cubic drift (Eq. \ref{eq:cubic_drift}) are not free parameters: they are state-dependent macroscopic quantities that must be recomputed in every computational cell at every time step. They are determined by enforcing the moment consistency condition (Eq. \ref{eq:moment_match}) for the second-order (stress tensor) and third-order (heat flux vector) moments, a procedure, detailed in Gorji et al.~\cite{gorji2011fokker,gorji2014efficient}, that generates a $9 \times 9$ system of linear equations. Arranging the unknowns into the vector
$Y = \big(c_{11}^{(1)},\, c_{12}^{(1)},\, c_{13}^{(1)},\, c_{22}^{(1)},\, c_{23}^{(1)},\, c_{33}^{(1)},\, c_{1}^{(2)},\, c_{2}^{(2)},\, c_{3}^{(2)}\big)^{\mathsf{T}}$,
the system reads in block form~\cite{jenny2010}:}
\begin{equation}
\begin{pmatrix} V_{6 \times 6} & W_{6 \times 3} \\ X^T_{3 \times 6} & Z_{3 \times 3} \end{pmatrix}
\begin{pmatrix} Y_6 \\ Y_3 \end{pmatrix} =
\begin{pmatrix} Q_6 \\ R_3 \end{pmatrix}
\label{eq:system9x9}
\end{equation}
The matrix blocks $V, W, X^T, Z$ are composed of various velocity moments 
$u^{(p)}_{i_1\ldots i_n} = \int\allowbreak |v'|^p v'_{i_1}\allowbreak \ldots v'_{i_n}\allowbreak f\, d^3V,$
which are computed from the particle ensemble in the cell. The right-hand side vector $(Q_6, R_3)$ is derived from the Boltzmann collision operator's moment relaxation rates. For the heat flux relaxation (the $R_3$ block), this is non-zero and corresponds to the right-hand side of the heat flux equation.

\subsection{System Matrix Components}

The explicit components of the $9 \times 9$ system matrix are essential for implementation. The primary blocks $V_{6 \times 6}$ and $W_{6 \times 3}$, are detailed in Table \ref{tab:matrix_system}; the notation follows the original cubic-FP derivations of Gorji et al.~\cite{gorji2011fokker,gorji2014efficient}.

\begin{table}[h!]
\centering
\caption{Matrix blocks for the $9 \times 9$ linear system, derived from Gorji et al. Entries are defined by velocity moments $u^{(p)}_{...} = \int |v'|^p v'_{i_1}... v'_{i_n} f d^3V$.}
\label{tab:matrix_system}
\begin{tabular}{@{}p{\textwidth}@{}}
\toprule
\textbf{Block $V_{6 \times 6}$ (Relating $c_{ij}^{(1)}$ to stress relaxation)} \\
\midrule
$
V_{6 \times 6} = \begin{pmatrix}
2u^{(2)}_{11} & 2u^{(2)}_{12} & 2u^{(2)}_{13} & 0 & 0 & 0 \\
u^{(2)}_{12} & u^{(2)}_{22} + u^{(2)}_{11} & u^{(2)}_{23} & u^{(2)}_{12} & u^{(2)}_{13} & 0 \\
u^{(2)}_{13} & u^{(2)}_{23} & u^{(2)}_{33} + u^{(2)}_{11} & 0 & u^{(2)}_{12} & u^{(2)}_{13} \\
0 & 2u^{(2)}_{12} & 0 & 2u^{(2)}_{22} & 2u^{(2)}_{23} & 0 \\
0 & u^{(2)}_{13} & u^{(2)}_{12} & u^{(2)}_{23} & u^{(2)}_{33} + u^{(2)}_{22} & u^{(2)}_{23} \\
0 & 0 & 2u^{(2)}_{13} & 0 & 2u^{(2)}_{23} & 2u^{(2)}_{33}
\end{pmatrix}
$
\\
\addlinespace[10pt]
\midrule
\textbf{Block $W_{6 \times 3}$ (Relating $c_{i}^{(2)}$ to stress relaxation)} \\
\midrule
$
W_{6 \times 3} = \begin{pmatrix}
2u^{(2)}_{1} & 0 & 0 \\
u^{(2)}_{2} & u^{(2)}_{1} & 0 \\
u^{(2)}_{3} & 0 & u^{(2)}_{1} \\
0 & 2u^{(2)}_{2} & 0 \\
0 & u^{(2)}_{3} & u^{(2)}_{2} \\
0 & 0 & 2u^{(2)}_{3}
\end{pmatrix}
$
\\
\bottomrule
\end{tabular}
\end{table}

\subsection{Particle-Based Numerical Implementation}

\subsubsection{The Equivalent Itô Stochastic Process}

The particle-based simulation algorithm does not solve the PDE (Eq. \ref{eq:fp_general}) directly. Instead, it solves the mathematically equivalent system of Itô stochastic differential equations (SDEs) that describe the evolution of a single particle's position $X(t)$ and velocity $M(t)$. This system is given by:
\begin{align}
dM_i &= A_i dt + d_{ij} dW_j \label{eq:sde_vel} \\
dX_i &= M_i dt \label{eq:sde_pos}
\end{align}
where $d_{ij}$ is the matrix square root of the diffusion tensor ($d_{ik} d_{jk} = D_{ij}$), and $dW_j$ is a standard three-dimensional Wiener process (a Gaussian random variable with mean 0 and variance $dt$). For the cubic-FP model with its simple isotropic diffusion (Eq. \ref{eq:cubic_diffusion}), the term $d_{ij} dW_j$ simplifies to $\sqrt{2\theta/\tau} \delta_{ij} dW_j$.

\subsection{Mixed Time Integration Scheme for Particle Velocity}

\rev{The SDE system (Eq. \ref{eq:sde_vel}-\ref{eq:sde_pos}) is non-linear and stiff, and a simple Euler-Maruyama scheme is insufficient. It is integrated with the mixed scheme of Gorji \& Jenny~\cite{gorji2011fokker}: the linear part of the drift ($-\frac{1}{\tau}v_i'$) is integrated exactly as an Ornstein--Uhlenbeck process, which stabilizes the stiff relaxation to equilibrium, while the remaining non-linear parts are advanced with a first-order Euler step.}

The resulting velocity update equation for a particle from time $t^n$ to $t^{n+1} = t^n + \Delta t$ is:
\begin{equation}
M_i^{n+1} = \frac{1}{\alpha}\left(M_i^{\prime n}e^{-\Delta t/\tau} + (1 - e^{-\Delta t/\tau})\tau N_i^n + \sqrt{\theta^n (1 - e^{-2\Delta t/\tau})}\xi_i\right) + U_i^n
\label{eq:update_vel}
\end{equation}
where:
\begin{itemize}
    \item $M_i^{\prime n} = M_i^n - U_i^n$ is the particle's peculiar velocity at time $t^n$.
    \item $U_i^n$ and $\theta^n$ are the cell's mean velocity and temperature, assumed constant over $\Delta t$.
    \item $\xi_i$ is a standard normal random variate (mean 0, variance 1) sampled from a Gaussian distribution.
    \item $N_i^n$ is the non-linear contribution to the drift.
    \item $\alpha$ is a numerical correction factor to enforce energy conservation.
\end{itemize}

\subsection{Nonlinear Contribution and Energy Conservation}

The non-linear contribution $N_i^n$ is simply the non-linear part of the cubic drift model (Eq. \ref{eq:cubic_drift}), evaluated using the particle's velocity $M'$ and the cell's macroscopic coefficients:
\begin{equation}
N_i^n = \left( c_{ij}^{(1)} + \frac{1}{\tau}\delta_{ij} \right) M_j^{\prime n} + c_i^{(2)}(M_j^{\prime n}M_j^{\prime n} - 3\theta^n) - \epsilon^2\left(M_i^{\prime n}M_j^{\prime n}M_j^{\prime n} - \frac{2q_i^n}{\rho^n}\right)
\label{eq:update_N}
\end{equation}
Note that the linear model's drift ($-\frac{1}{\tau}M_j^{\prime n}$) has been added back to $c_{ij}^{(1)}$ because the "exact" integration part only accounted for $M_i^{\prime n}e^{-\Delta t/\tau}$.

\rev{The factor $\alpha$ is a non-physical numerical correction that repairs the energy-conservation error introduced by the first-order Euler integration of the $N_i^n$ term: the particle velocity is rescaled by $1/\alpha$ at the end of the step so that the expected energy of the particle ensemble is conserved. The factor is computed from its squared value:}
\begin{equation}
\alpha^2 = 1 + \frac{\tau}{3\theta}\left(\tau(1 - e^{-\Delta t/\tau})^2 \langle N_i^n N_i^n \rangle + 2(e^{-\Delta t/\tau} - e^{-2\Delta t/\tau})\langle M_i^{\prime n}N_i^n \rangle\right)
\label{eq:update_alpha}
\end{equation}
The angle brackets $\langle \dots \rangle$ denote the ensemble average (i.e., the mean) computed over all particles within the computational cell.

\subsection{Position Update}

The position update is typically handled with a simple and explicit first-order Euler integration, which is sufficiently accurate as the position SDE has no stochastic term:
\begin{equation}
X_i^{n+1} = X_i^n + M_i^n \Delta t
\label{eq:update_pos}
\end{equation}

\section{Numerical Methodology: A GPU-Native Surrogate}

\begin{figure}[t]
\centering
\includegraphics[width=0.98\textwidth]{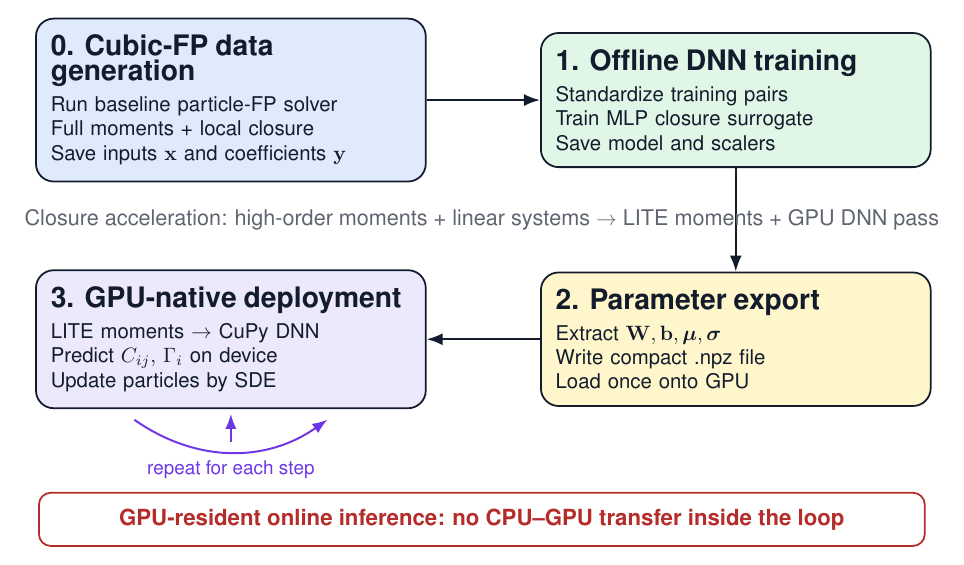}
\caption{GPU-native surrogate pipeline. The offline stage generates cubic-FP training pairs, trains the neural surrogate, and exports weights and normalization parameters. During online deployment, the exported arrays are loaded once onto the GPU, and the particle simulation loop uses low-order LITE moments plus a pure-CuPy forward pass to obtain closure coefficients without CPU--GPU transfers.}
\label{fig:flowchart}
\end{figure}

Our approach consists of a four-phase pipeline, as illustrated in Figure \ref{fig:flowchart}. We first build a baseline physics solver (Phase 0) to generate data, then train a surrogate (Phase 1-2), and finally deploy it back into a simulation (Phase 3).

\subsection{Phase 0: The Baseline Physics Solver}
To establish a high-performance baseline, the entire simulation was implemented in Python using the \texttt{CuPy} library, which provides a CUDA-accelerated, NumPy-like API. This allows all particle and grid data to remain in GPU memory. The simulation loop is detailed in Algorithm \ref{alg:baseline}.

The critical steps are data-intensive moment gathering and compute-intensive linear solve. These two steps constitute the "Physics Solver" component, which we target for replacement.

\begin{algorithm}[h!]
\caption{Baseline Physics Simulation Loop}
\label{alg:baseline}
\begin{algorithmic}
\State Initialize $N_p$ particles on GPU (\texttt{p\_data})
\State Initialize $N_c$ grid cells on GPU (\texttt{grid\_gpu}, \texttt{coeffs\_gpu}, \texttt{linsys\_gpu})
\For{$nt = 1$ to $N_{\text{steps}}$}
    \State \texttt{Move\_Particles\_2D}(\texttt{p\_data}, \texttt{DT})
    \State \texttt{Apply\_Boundary\_Cavity}(\texttt{p\_data})
    \State \texttt{sort\_and\_calc\_moments\_FULL}(\texttt{p\_data}, \texttt{grid\_gpu}) \label{alg:step:full_moments} \Comment{Calculates M1-M5}
    \If{$nt > N_{\text{ss}}$}
        \State \texttt{average\_results}(\texttt{avg\_grid\_gpu}, \texttt{grid\_gpu}, $nt$, $N_{\text{ss}}$)
    \EndIf
    \State \texttt{build\_linear\_systems}(\texttt{grid\_gpu}, \texttt{linsys\_gpu}) \label{alg:step:solver_build}
    \State \texttt{solve\_linear\_systems}(\texttt{linsys\_gpu}, \texttt{coeffs\_gpu}) \label{alg:step:solver_solve} \Comment{\textbf{Target Bottleneck}}
    \State \texttt{evolve\_velocities}(\texttt{p\_data}, \texttt{grid\_gpu}, \texttt{coeffs\_gpu}, \texttt{DT})
\EndFor
\State Transfer \texttt{avg\_grid\_gpu} to CPU for plotting.
\end{algorithmic}
\end{algorithm}

\subsection{Phase 1-2: Data Generation and DNN Training}
\label{sec:phase12}
We first generated a training dataset from the 1D Couette flow problem, running 20 simulations with randomized boundary conditions. At each sampling step, we saved the inputs and outputs of the physics solver.

1. Input Features ($\mathbf{X} \in \mathbb{R}^{16}$): We selected 16 low-order moments and properties:
    $\rho, T, U_x, U_y, U_z, \Pi_{xx}, \Pi_{xy}, \Pi_{xz}, \Pi_{yy}, \Pi_{yz}, \Pi_{zz}, q_x, q_y, q_z, DM_2, \nu$.

2. Target labels ($\mathbf{Y} \in \mathbb{R}^{9}$): the six independent components of $C_{ij}$ and the three components of $\Gamma_i$. These correspond to $c^{(1)}_{ij}$ and $c^{(2)}_i$ in the cubic-FP derivation.

This dataset was used to train a 4-layer MLP using Keras. The model \texttt{fp\_model.keras} and scalers \texttt{scaler\_X.pkl}, \texttt{scaler\_y.pkl} were saved.

\subsection{Phase 3: GPU-Native Deployment}
\label{sec:phase3}
This is the most critical phase for performance. A naive implementation that calls \texttt{model.predict()} in the simulation loop is unacceptably slow due to CPU-GPU data transfer. We therefore developed a GPU-native deployment strategy.

1. Parameter Extraction: A one-time script (\texttt{extract\_params.py}) was run to load the Keras/SKLearn files and dump all parameters---weights ($\mathbf{W}$), biases ($\mathbf{b}$), means ($\bm{\mu}$), and scales ($\bm{\sigma}$)---into a single NumPy \texttt{.npz} file (\texttt{model\_params\_for\_cupy.npz}).

2. Optimization (LITE Moments): We created a new moment function, \texttt{...LITE}, which skips the expensive \texttt{bincount} operations for $M_3, M_4, M_5, \lambda$, etc., and only computes the 16 features required by the ML model.

3. GPU-Native Solver: At the start of the new simulation, the \texttt{.npz} file is loaded and all parameters are transferred to GPU memory as \texttt{CuPy} arrays (e.g., \texttt{W1\_gpu}, \texttt{b1\_gpu}, \texttt{X\_mean\_gpu}). We then replace the physics solver (Steps \ref{alg:step:solver_build}-\ref{alg:step:solver_solve}) with a single function call, \texttt{predict\_coeffs\_cupy\_native}, which executes the entire DNN forward pass on the GPU.

The resulting accelerated simulation loop is shown in Algorithm \ref{alg:ml}.

\begin{algorithm}[h!]
\caption{Accelerated ML Simulation Loop}
\label{alg:ml}
\begin{algorithmic}
\State Initialize $N_p$ particles on GPU (\texttt{p\_data})
\State Initialize $N_c$ grid cells on GPU (\texttt{grid\_gpu}, \texttt{coeffs\_gpu})
\State \texttt{Load\_ML\_Params\_to\_GPU}(\texttt{model\_params.npz}) $\rightarrow$ \texttt{ml\_params\_gpu}
\For{$nt = 1$ to $N_{\text{steps}}$}
    \State \texttt{Move\_Particles\_2D}(\texttt{p\_data}, \texttt{DT})
    \State \texttt{Apply\_Boundary\_Cavity}(\texttt{p\_data})
    \State \texttt{sort\_and\_calc\_moments\_LITE}(\texttt{p\_data}, \texttt{grid\_gpu}) \label{alg:step:lite_moments} \Comment{Calculates M1-M3 (density, velocity, stress, heat flux) required by DNN}
    \If{$nt > N_{\text{ss}}$}
        \State \texttt{average\_results}(\texttt{avg\_grid\_gpu}, \texttt{grid\_gpu}, $nt$, $N_{\text{ss}}$)
    \EndIf
    \State \texttt{predict\_coeffs\_cupy\_native}(\texttt{grid\_gpu}, \texttt{coeffs\_gpu}, \texttt{ml\_params\_gpu}) \label{alg:step:ml_solve} \Comment{\textbf{New Solver}}
    \State \texttt{evolve\_velocities}(\texttt{p\_data}, \texttt{grid\_gpu}, \texttt{coeffs\_gpu}, \texttt{DT})
\EndFor
\State Transfer \texttt{avg\_grid\_gpu} to CPU for plotting.
\end{algorithmic}
\end{algorithm}

The performance of Algorithm \ref{alg:baseline} (Physics) versus Algorithm \ref{alg:ml} (ML) is the central comparison of this study.

\section{Results and Discussion}

We validate our approach on three canonical flow problems. \rev{All production simulations and profiling runs were performed on NVIDIA GPUs available through the UMass Amherst Unity cluster, using the 48-GB and 80-GB VRAM GPU partitions when available. Auxiliary debugging and short verification runs were also carried out on other NVIDIA GPUs, including Tesla M40 24-GB and A16 devices; therefore, the reported wall-clock timings should be interpreted as hardware-specific benchmarks for the stated GPU class rather than universal performance numbers.}

\subsection{Case Study 1: 1D Couette Flow}
The 1D Couette flow serves as the first validation case. The working gas for all simulations was Argon ($m_{Ar} = \SI{66.3e-27}{\kilo\gram}$), modeled as a Maxwell molecule ($\omega=1.0$) with a reference viscosity $\mu_0 = \SI{2.117e-5}{\pascal\second}$ and a specific heat ratio $\gamma = 5/3$. The baseline parameters, including an initial gas temperature ($T_{in}$) and wall temperature ($T_{w}$) of \SI{273.15}{\kelvin}, were set to establish a baseline Knudsen number of $Kn \approx 0.15$.

\rev{The complete simulation specification for this benchmark (gas properties, flow and boundary conditions, discretization, particle counts, and run lengths) is provided in Supplementary Material Table~S1; the key settings of the profile-verification run are a $L_x=1$~mm domain with $N_C=300$ cells, $N_P = 9{,}000{,}000$ particles (30{,}000 per cell), opposing wall velocities of $\pm 50$~m/s at $T_w=273.15$~K, and a 30{,}000-step run with time averaging after 10{,}000 steps.}

\subsubsection{Accuracy Validation}
\begin{figure}[htbp!]
    \centering
    \begin{subfigure}{0.48\textwidth}
        \centering
        \includegraphics[width=\linewidth]{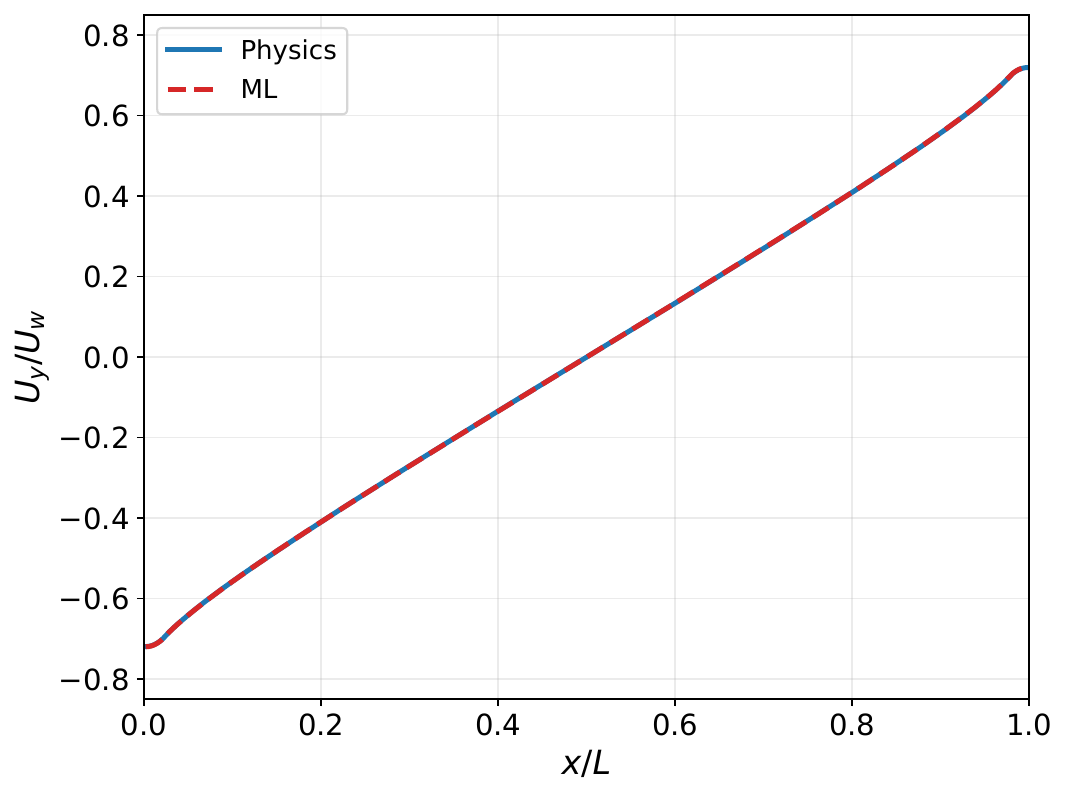}
        \caption{Velocity.}
    \end{subfigure}\hfill
    \begin{subfigure}{0.48\textwidth}
        \centering
        \includegraphics[width=\linewidth]{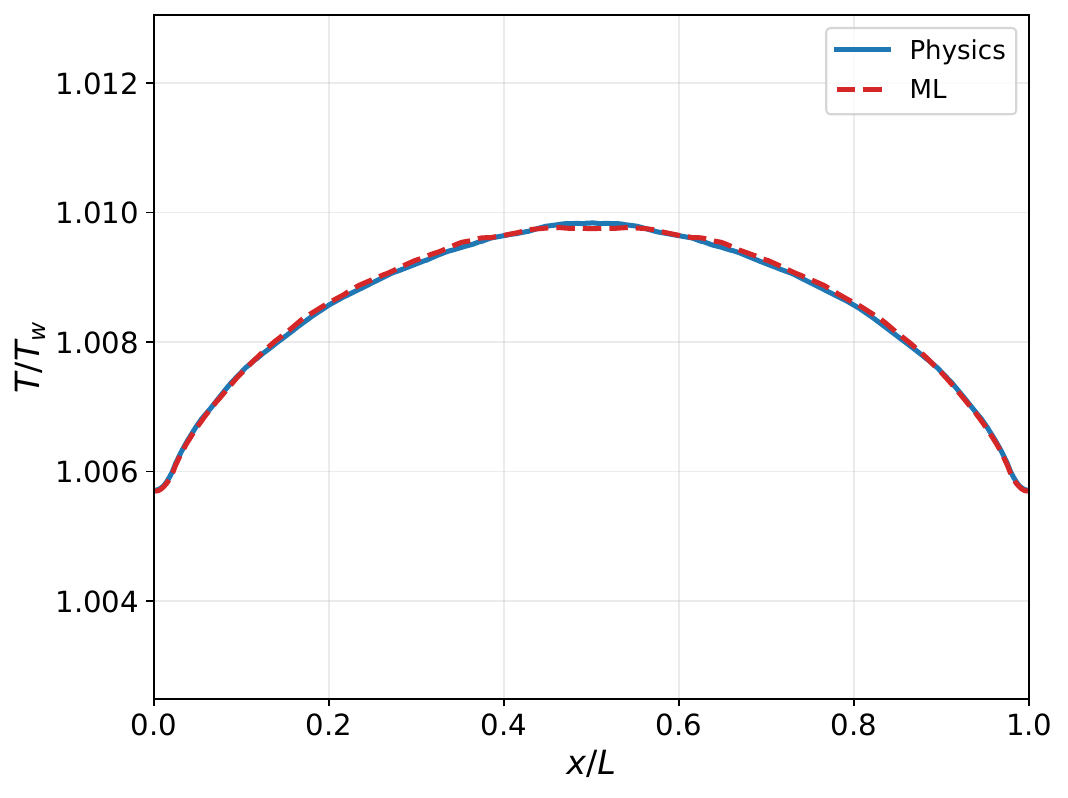}
        \caption{Temperature.}
    \end{subfigure}
    \caption{Compact Couette verification at nominal $Kn\approx0.15$. Profiles are nondimensionalized as $U_y/U_w$ and $T/T_w$. Solid curves denote the cubic-FP physics baseline and dashed curves denote the GPU-native ML surrogate.}
    \label{fig:couette_comparison}
\end{figure}

Figure~\ref{fig:couette_comparison} compares the final time-averaged, dimensionless velocity and temperature profiles from the cubic-FP physics baseline and the GPU-native ML surrogate. The ML surrogate reproduces the expected slip-modified velocity profile and the symmetric viscous-heating temperature hump, confirming that the deployed closure produces the expected one-dimensional behavior before the multidimensional cavity and cylinder tests.

\subsubsection{Performance Benchmark}
Table~\ref{tab:couette_performance} reports a compact timing run ($N_c=100$, $N_p=3$M, 4000 steps), separate from the longer profile-averaging run used in Fig.~\ref{fig:couette_comparison}. The ML-driven simulation replaces the full high-order moment assembly and dense local solve by a low-order moment pass plus native GPU inference, giving a 1.56$\times$ speedup. The measured net online saving is $262.12-167.73=94.39$ s for this timing configuration.

\begin{table}[h!]
  \centering
  \caption{Performance: 1D Couette Flow (4000 steps, $N_c=100$, $N_p=3\text{M}$)}
  \label{tab:couette_performance}
  \begin{tabularx}{\textwidth}{l >{\raggedright\arraybackslash}X S[table-format=3.2] r}
    \toprule
    \textbf{Solver} &
    \textbf{Solver Method} &
    \textbf{Time (s)} &
    \textbf{Speedup} \\
    \midrule
    Physics Baseline &
      \makecell[l]{\texttt{...\_FULL} \\ + \texttt{cp.linalg.solve}} &
      262.12 & 1.0x \\
    \textbf{ML} &
      \makecell[l]{\texttt{...\_LITE} \\ + \texttt{predict\_native}} &
      \textbf{167.73} & \textbf{1.56x} \\
    \bottomrule
  \end{tabularx}
\end{table}

\subsubsection{Extension to a Wide Range of Knudsen Numbers}
In the next stage, the aim is to train our DNN over a wide range of Knudsen number for the one-dimensional planar Couette flow problem. \rev{To this end, the baseline case was rescaled in density by the factors \texttt{kn\_factors = [0.01, 0.1, 0.5, 1.0, 2.0]}, producing five training simulations at nominal $Kn \approx 0.0015$, $0.015$, and $0.075$ (slip regime) and $Kn \approx 0.15$ and $0.3$ (transition regime). All runs used $N_C = 300$ cells and $N_P = 9{,}000{,}000$ particles (30{,}000 per cell) to suppress statistical noise. Because the high-$Kn$ cases approach steady state slowly, every simulation was run for $40{,}000$ steps with a conservative $30{,}000$-step transient, and only the final $10{,}000$ steady-state steps were sampled. At each sampled step and cell, the 16 low-order input features defined in Section~\ref{sec:phase12} and the corresponding nine cubic-FP closure coefficients ($C_{ij}$, $\Gamma_i$) returned by the exact $9 \times 9$ solve were stored, yielding $10{,}000 \times 300 \times 5 = 15{,}000{,}000$ training pairs that map the local macroscopic state directly to the required closure coefficients. Further data-generation and convergence details are given in the Supplementary Material (Section~S2).}

\subsubsection{Neural Network Architecture}

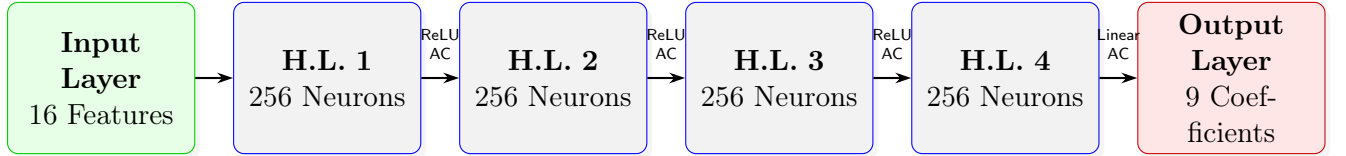
\begin{figure}[h!]
  \centering
  
  \tikzstyle{layer} = [
    rectangle, 
    rounded corners, 
    draw=blue, 
    fill=gray!10, 
    minimum height=2.0cm, 
    minimum width=2.4cm,  
    text width=2.2cm,     
    align=center,
    font=\small,          
    drop shadow={opacity=0.4, fill=gray!20, shadow xshift=2pt, shadow yshift=-2pt}
  ]
  \tikzstyle{arrow} = [draw, -{Stealth[length=6pt, width=4pt]}, thick, black] 
  
  \tikzstyle{annote} = [
    text=black, 
    font=\tiny\sffamily, 
    midway, 
    above, 
    align=center,
    yshift=0.1cm 
  ] 

  \begin{tikzpicture}[node distance=1.5cm and 0.5cm] 

    \node (input) [layer, fill=green!10, draw=green!80!black] 
      {\textbf{Input Layer} \\ 16 Features};
      
    \node (hl1) [layer, right=of input] 
      {\textbf{H.L. 1} \\ 256 Neurons};
      
    \node (hl2) [layer, right=of hl1] 
      {\textbf{H.L. 2} \\ 256 Neurons};
      
    \node (hl3) [layer, right=of hl2] 
      {\textbf{H.L. 3} \\ 256 Neurons};
      
    \node (hl4) [layer, right=of hl3] 
      {\textbf{H.L. 4} \\ 256 Neurons};

    \node (output) [layer, fill=red!10, draw=red!80!black, right=of hl4] 
      {\textbf{Output Layer} \\ 9 Coefficients};

    \draw [arrow] (input) -- (hl1);
    \draw [arrow] (hl1)   -- (hl2)   node [annote] {ReLU \\ AC};
    \draw [arrow] (hl2)   -- (hl3)   node [annote] {ReLU \\ AC};
    \draw [arrow] (hl3)   -- (hl4)   node [annote] {ReLU \\ AC};
    \draw [arrow] (hl4)   -- (output) node [annote] {Linear \\ AC};

  \end{tikzpicture}
  
  \caption{Schematic of the 16-256-256-256-256-9 feed-forward Deep Neural Network (DNN) architecture used as the surrogate model. "H.L." stands for Hidden Layer. "AC" stands for Activation.}
  \label{fig:nn_schematic}
\end{figure}

The surrogate model is a fully-connected, feed-forward Deep Neural Network (DNN), also known as a Multi-Layer Perceptron (MLP). \rev{Its architecture, shown in Figure~\ref{fig:nn_schematic} and summarized as 16-256-256-256-256-9, consists of four ReLU hidden layers of 256 neurons each between a 16-feature input layer and a 9-coefficient linear output layer (a regression head), for a total of 204{,}041 trainable parameters.}

\par
\rev{This configuration was selected in a preliminary hyperparameter study: shallower networks (1--2 hidden layers) underfit the strongly non-linear 16-to-9 moment-to-coefficient map, while substantially wider (e.g., 1024-neuron) or deeper networks increased the online inference cost that the method is designed to minimize without improving the model-selection loss ($< 2 \times 10^{-4}$); a more exhaustive search was therefore not pursued.}

\par
\rev{The 16 input features are exactly the low-order moments and properties listed in Section~\ref{sec:phase12}, all of which are computed online by the LITE moment pass; no moment above third order enters the network. This choice is deliberate: it is what allows the deployed surrogate to bypass entirely the high-order moment gathering ($M_3$, $M_4$, $M_5$) that dominates the cost of the deterministic closure.}

\rev{The deployed model is not the backpropagation-enabled Keras training object: as described in Section~\ref{sec:phase3}, the trained weights, biases, and normalization scalers are exported once and evaluated as a forward-pass-only sequence of batched, GPU-native \texttt{CuPy} matrix operations inside the simulation loop.}

\subsubsection{Training and Validation}

The pooled 15-million-sample dataset was randomly partitioned only for optimizer monitoring, early stopping, and checkpoint selection. This cell-level split is not used as the evidence for physical generalization because neighboring cells and adjacent time samples are correlated. Generalization is instead assessed by complete held-out simulation conditions, as summarized in Table~\ref{tab:cavity_train_test_conditions}. The word ``validation'' below therefore refers to model-selection loss unless otherwise stated.

\rev{Inputs and outputs were standardized with a Scikit-learn \texttt{StandardScaler} fitted on the training split only and then applied to both splits. The network was trained with the Adam optimizer (learning rate $1 \times 10^{-4}$) on a mean-squared-error loss, using an \texttt{EarlyStopping} callback (patience of 10 epochs on the model-selection loss) and a \texttt{ModelCheckpoint} callback retaining the best model. The random cell-level validation loss was used only for early stopping and model selection; the final accuracy claims in this paper are based on complete simulation-condition tests rather than on this random split.}


\subsubsection{Robustness to Flow Regime (Knudsen Sweep)}

To test the performance gains across different flow regimes, we performed a parameter sweep on the 1D Couette flow problem. The baseline Knudsen number (Kn) of $\approx$0.15 (transitional flow) was varied over three orders of magnitude, from the near-continuum regime (Kn $\approx$0.0015) to the mid-transitional regime (Kn $\approx$1.5). This was achieved by decreasing (or increasing) the base density ($\rho$), while all other parameters (geometry, Np=9M, Nc=300) were held constant.

The objective was to determine if the computational cost of the physics solver is sensitive to the collision frequency $\nu$ (which is proportional to $\rho$), and if the ML model's speedup is maintained in these different physical contexts.

\subsubsection{Error Analysis}
Table \ref{tab:perf_comparison} summarizes the predictive fidelity of the Couette surrogate across representative Knudsen numbers. This table is intentionally separated from the timing sweep in Table~\ref{tab:kn_sweep}: Table~\ref{tab:perf_comparison} reports accuracy for selected validation profiles, whereas Table~\ref{tab:kn_sweep} reports a separate timing sweep with fixed $N_c=300$, $N_p=9$M, and 4000 steps. Within the training range ($Kn \in [0.0015,0.3]$), the relative temperature error remains below $0.07\%$. In the $Kn=0.7$ extrapolation case, the velocity-profile error increases, as expected, but the run remains stable. These results support robustness for the one-dimensional Couette manifold only and are not used to claim unrestricted transferability.

\begin{table}[htbp]
    \centering
    \caption{Representative Couette profile errors across Knudsen number. The ML model was trained on a regime of $Kn \in [0.0015,0.3]$. These are accuracy checks; timing for the fixed-grid sweep is reported separately in Table~\ref{tab:kn_sweep}.}
    \label{tab:perf_comparison}
    \renewcommand{\arraystretch}{1.2}
    \begin{tabular}{lccc}
        \hline
        \textbf{Knudsen No.} & \textbf{$L_2$ Error ($U_y$)} & \textbf{Rel. Error ($T$)} & \textbf{Regime} \\
        \hline
        $Kn=0.05$ & $5.19 \times 10^{-3}$ & $0.06\%$ & Interpolation \\
        $Kn=0.09$ & $4.11 \times 10^{-3}$ & $0.03\%$ & Interpolation \\
        $Kn=0.7$  & $2.30 \times 10^{-2}$ & $0.06\%$ & Extrapolation \\
        \hline
    \end{tabular}
\end{table}

\subsubsection{Performance Analysis}
Table \ref{tab:kn_sweep} summarizes the performance of both solvers across the four tested Knudsen numbers. The results are highly insightful and confirm a key aspect of this computational model:

1. Constant Physics Time: The runtime of the \texttt{Physics Baseline} solver remained nearly constant at $\approx 448\pm5$ seconds, regardless of the 1,000-fold change in Knudsen number. This indicates that, for this fixed grid and particle count, the computational cost of the physics solver (calculating M3-M5 and solving 300×(9×9) systems) is dominated by the number of operations rather than by the particular values of $\nu$ or $\rho$.

2. Constant ML Time: The runtime of the \texttt{ML} solver also remained nearly constant ($\approx 270 \pm 15$ seconds). This is expected, as its runtime is dominated by the $T_{\text{particles\_LITE}}$ component, which depends on the total number of particles ($N_p = 9M$), not the physical properties.

3. Consistent Speedup: As a result, the speedup achieved by the ML surrogate remained stable and significant across the entire 1,000-fold range of Knudsen numbers, consistently performing $\approx 1.63\text{x}$ to $\approx 1.73\text{x}$ faster than the physics solver.

\begin{table}[h!] \centering \caption{Performance Comparison for 1D Couette Flow Across Varying Knudsen Numbers (4000 steps, Nc=300, Np=9M)} \label{tab:kn_sweep} \begin{tabularx}{\textwidth}{l r r r} \toprule \textbf{Knudsen No. (Approx.)} & \textbf{$\mathrm{\textbf{Time\textsubscript{Physics} (s)}
}$ 
} & \textbf{$\mathrm{\textbf{Time\textsubscript{ML} (s)}
}$ 
} & \textbf{Speedup} \\ \midrule Kn $\approx$ 0.0015 & 448.89 s & 266.96 s & 1.68x \\ Kn $\approx$ 0.015 & 448.27 s & 259.26 s & 1.73x \\ Kn $\approx$ 0.15 & 449.73 s & 269.41 s & 1.67x \\ Kn $\approx$ 1.50 & 449.24 s & 275.09 s & 1.63x \\ \bottomrule \end{tabularx} \end{table}

This test indicates that the ML surrogate provides a consistent performance benefit across the tested Couette densities. The result should be interpreted as evidence of robustness for this one-dimensional benchmark, not as proof of unrestricted transferability to arbitrary rarefied flows.


\subsection{Case Study 2: 2D Lid-Driven Cavity}
\label{sec:cavity_revised}

\rev{The lid-driven cavity is the main internal-flow validation problem of this work. Unlike the preliminary Couette checks, the cavity contains two-dimensional recirculation, strong shear near the moving wall, compression and thermal gradients near the corners, and non-trivial nonequilibrium structure. It therefore provides a more relevant test of whether the GPU-native surrogate can replace the deterministic cubic-FP closure in a multidimensional particle simulation.}

\subsubsection{Training protocol and identifiability limitation}
\rev{To avoid optimistic estimates caused by correlated neighboring cells and successive time steps, the cavity surrogate is evaluated by complete simulation conditions. The training data were generated from representative cavity simulations over a range of lid velocities, and the validation runs reported below are performed on unseen target simulations. The surrogate maps a 16-component low-order moment vector to the cubic-FP closure coefficients. This map is not claimed to be universally identifiable in the full kinetic space: two different velocity distribution functions can share the same low-order moments but differ in higher-order structure. The learned closure should therefore be interpreted as a data-driven approximation valid on the distribution manifold represented by the training simulations, not as a universal kinetic closure.}

\begin{table}[t]
\centering
\caption{Cavity training and held-out test conditions for the q-weighted surrogate. The random cell-level split is used only for early stopping; the evidence for generalization is taken from complete simulations not used in fitting.}
\label{tab:cavity_train_test_conditions}
\footnotesize
\setlength{\tabcolsep}{4pt}
\renewcommand{\arraystretch}{1.12}
\resizebox{\textwidth}{!}{%
\begin{tabular}{lcccl}
\hline
Set & Nominal $Kn$ & $U_{\rm lid}$ (m/s) & Time steps / samples & Role \\
\hline
Training runs & 0.15 & 50, 100, 200, 400, 600 & 35,000 steps each; 15M pooled samples & Fit DNN weights \\
Cell-level split & 0.15 & Same as training runs & 90/10 random split & Early stopping only \\
Held-out speed test & 0.15 & 800 & Complete simulation & Unseen lid velocity \\
Held-out rarefaction test & 0.5 & 800 & Complete simulation & Rarefaction stress test \\
Held-out rarefaction test & 1.0 & 800 & Complete simulation & Primary rarefied cavity figure \\
\hline
\end{tabular}%
}
\end{table}

\rev{The 2D cavity surrogate was trained at a nominal Knudsen number of $Kn=0.15$, using lid velocities $U_{\rm lid}=50$, 100, 200, 400, and 600 m/s. The $Kn=0.5$ and $Kn=1.0$ cases reported below are therefore not additional training conditions; they are held-out rarefaction stress tests at $U_{\rm lid}=800$ m/s. The increase in the high-order diagnostic errors, as we will see in the next sections, from $Kn=0.5$ to $Kn=1.0$ is thus expected: these cases move progressively farther from the nominal rarefaction level used during training and probe weaker overlap between the local distribution manifold seen by the network and the target particle distributions.}

\rev{In this work, ``q-weighted'' denotes the cavity training loss in which the heat-flux coefficient block $\Gamma_i$ is assigned a larger loss weight than the stress-relaxation block $C_{ij}$, so that the network is penalized more strongly for errors in the heat-flux channel. This weighting is used only during training; the deployed GPU-native inference evaluates the raw network outputs with no post-training projection.}

\rev{We also considered a deliberately difficult test in which a surrogate trained only on 1D Couette data was inserted into the 2D cavity solver. That case is discussed only briefly because it mainly demonstrates a negative result: representative training data are required. The 1D-trained model captured some primary velocity features but was not adequate for the full 2D thermal and nonequilibrium structure. The remainder of this section therefore focuses on the 2D-trained, q-weighted cavity surrogate.}

\subsubsection{Macroscopic-field validation at nominal \texorpdfstring{$Kn=1.0$}{Kn=1.0}}
\begin{landscape}
\begin{figure}[p]
    \centering
    \includegraphics[width=1.05\linewidth,height=0.80\textheight,keepaspectratio]{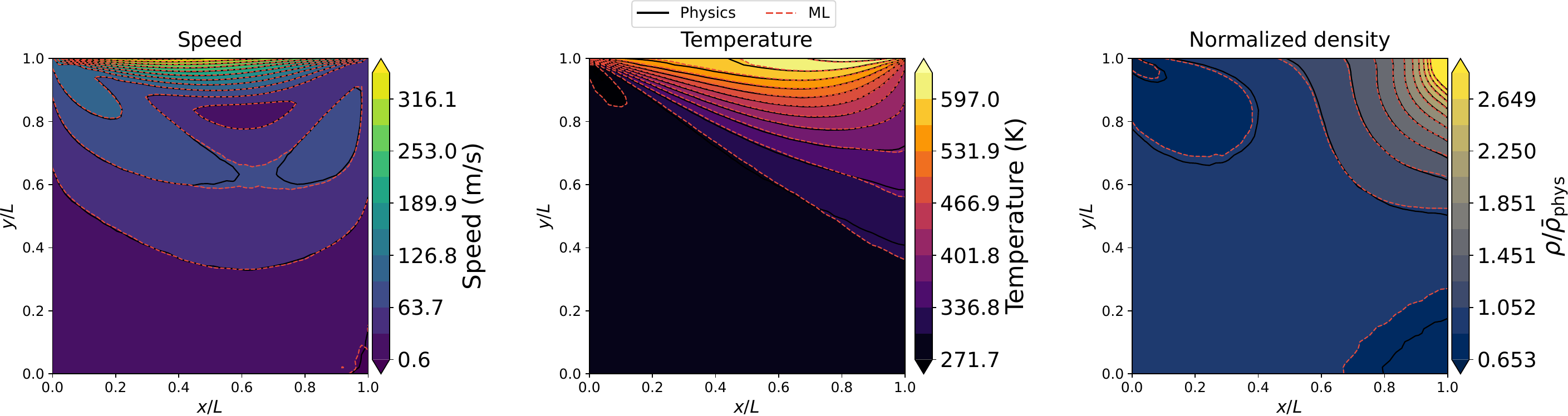}
    \caption{2D cavity validation at nominal $Kn=1.0$ and $U_{\rm lid}=800$ m/s. Filled contours are the ML solution, black solid lines are the cubic-FP physics baseline, and red dashed lines are ML contour lines. The axes are nondimensionalized by the cavity length.}
    \label{fig:cavity_macro_kn1}
\end{figure}
\end{landscape}

\rev{To probe higher rarefaction while remaining within the intended regime of the cubic-FP reference model, we performed cavity tests at nominal $Kn=0.5$ and $Kn=1.0$. The $Kn=1.0$ case is reported as the primary rarefied cavity figure because it is the more stringent of the two while avoiding the near-free-molecular limit. The axes are nondimensionalized by the cavity length. Figure~\ref{fig:cavity_macro_kn1} compares speed, temperature, and normalized density. The ML contours reproduce the physics baseline throughout the main vortex and the upper-wall thermal layer.}

\clearpage
\clearpage

\begin{table}[t]
\centering
\caption{Macroscopic-field errors for the q-weighted cavity surrogate.}
\label{tab:qweighted_macro_errors}
\footnotesize
\setlength{\tabcolsep}{4pt}
\renewcommand{\arraystretch}{1.12}
\resizebox{\textwidth}{!}{%
\begin{tabular}{lcccc}
\hline
Quantity & Relative $L_2$ error (\%) & Scaled RMSE (\%) & Max. scaled abs. error (\%) & Mean ML/Physics \\
\hline
$|\mathbf{U}|$ & 0.278 & 0.061 & 0.274 & 0.9996 \\
$T$ & 0.102 & 0.054 & 0.259 & 0.9999 \\
$\rho/\rho_0$ & 0.132 & 0.018 & 0.077 & 1.0000 \\
$p/p_0$ & 0.124 & 0.012 & 0.062 & 0.9998 \\
\hline
\end{tabular}%
}
\end{table}

\subsubsection{High-order nonequilibrium diagnostics}
\rev{Quantitative error information is reported for the key contour figures, together with normalized high-order diagnostics extracted from the final averaged particle fields. These diagnostics are not used as direct network inputs; they test whether the evolved ML particle distribution retains the same nonequilibrium structure as the physics baseline.}

\begin{landscape}
\begin{figure}[p]
    \centering
    \includegraphics[width=1.05\linewidth,height=0.80\textheight,keepaspectratio]{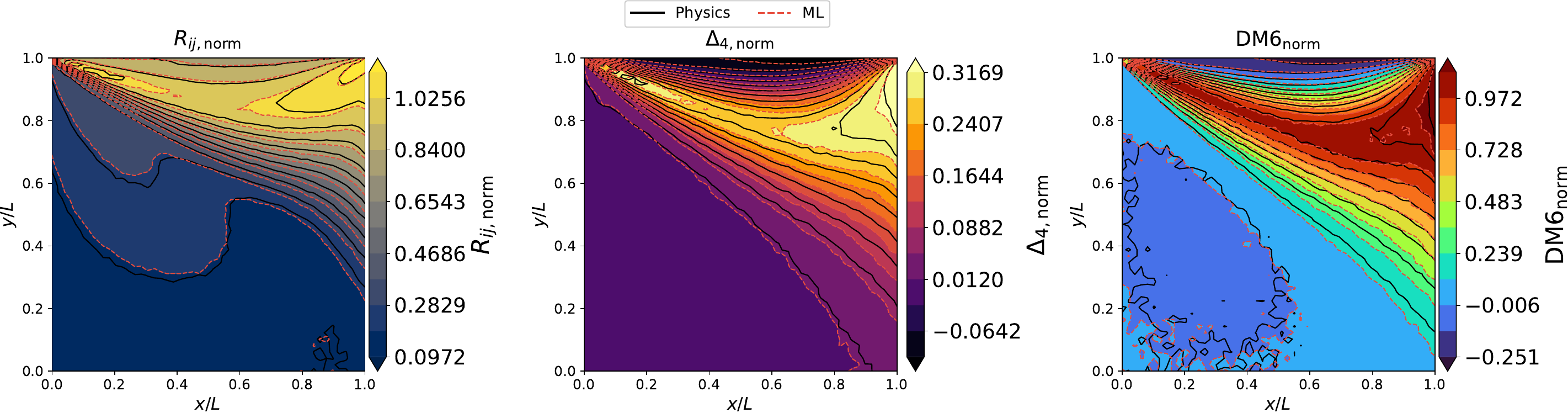}
    \caption{Normalized high-order nonequilibrium diagnostics for the rarefied cavity at nominal $Kn=1.0$. Filled contours are ML predictions, black solid contours are the physics baseline, and red dashed contours are ML contour lines.}
    \label{fig:cavity_highmom_kn1}
\end{figure}
\end{landscape}

\begin{landscape}
\begin{figure}[p]
    \centering
    \includegraphics[width=1.05\linewidth,height=0.80\textheight,keepaspectratio]{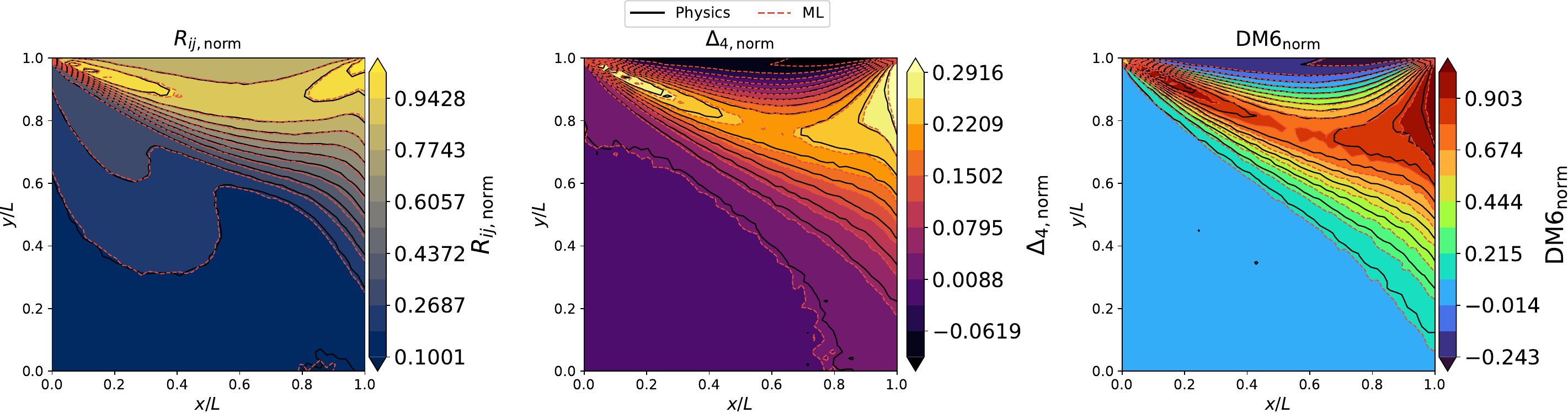}
    \caption{Normalized high-order nonequilibrium diagnostics for the rarefied cavity at nominal $Kn=0.5$. This case is included to demonstrate that the $Kn=1.0$ trends are not isolated to a single rarefaction level.}
    \label{fig:cavity_highmom_kn05}
\end{figure}
\end{landscape}

\rev{The high-order diagnostics shown in Fig.~\ref{fig:cavity_highmom_kn1} are computed from the peculiar velocity $c_i=v_i-U_i$. Here $R_{ij}=\langle c_i c_j c_k c_k\rangle$ is the contracted fourth-order tensor, $\Delta_4=\langle(c_k c_k)^2\rangle-15\theta^2$ is the fourth-order excess relative to a local Maxwellian, and $\mathrm{DM6}=\langle(c_k c_k)^3\rangle$ is the sixth central moment, with $\theta=\langle c_k c_k\rangle/3$. The subscript ``norm'' denotes nondimensionalization by the corresponding local Maxwellian scales, so that $R_{ij,\mathrm{norm}}$, $\Delta_{4,\mathrm{norm}}$, and $\mathrm{DM6}_{\mathrm{norm}}$ emphasize high-order nonequilibrium structure rather than density or temperature magnitude. Figure~\ref{fig:cavity_highmom_kn1} shows the $Kn=1.0$ comparison for these normalized diagnostics. The largest discrepancies appear in the top-wall shear layer and the upper corners, where the distribution is farthest from equilibrium. Nevertheless, the spatial organization of the high-order fields is preserved. Figure~\ref{fig:cavity_highmom_kn05} shows the corresponding $Kn=0.5$ case. Because the $Kn=0.5$ and $Kn=1.0$ contour trends are qualitatively similar, the main discussion emphasizes $Kn=1.0$, while the combined error table documents both cases.}

\clearpage
\clearpage

\clearpage
\clearpage

\begin{table}[t]
\centering
\caption{High-order nonequilibrium diagnostic errors for the rarefied cavity tests. The relative error is the global $L_2$ error, while RMSE is reported as $100\,\sqrt{\langle(\mathrm{ML}-\mathrm{Physics})^2\rangle}$ for the nondimensional normalized fields.}
\label{tab:cavity_kn_highmom_errors}
\setlength{\tabcolsep}{5pt}
\renewcommand{\arraystretch}{1.15}
\begin{tabular}{lcccc}
\toprule
Quantity & \multicolumn{2}{c}{Relative $L_2$ error (\%)} & \multicolumn{2}{c}{RMSE$\times 100$} \\
\cmidrule(lr){2-3}\cmidrule(lr){4-5}
 & Kn=0.5 & Kn=1.0 & Kn=0.5 & Kn=1.0 \\
\midrule
$\sigma_{\mathrm{norm}}$ & 0.728 & 1.008 & 0.133 & 0.198 \\
$q_{\mathrm{norm}}$ & 3.669 & 7.155 & 0.708 & 1.526 \\
$m_{3,\mathrm{norm}}$ & 1.650 & 2.751 & 0.250 & 0.464 \\
$R_{ij,\mathrm{norm}}$ & 1.566 & 3.337 & 0.706 & 1.642 \\
$\Delta_{4,\mathrm{norm}}$ & 5.610 & 10.089 & 0.574 & 1.164 \\
$\mathrm{DM6}_{\mathrm{norm}}$ & 6.904 & 13.642 & 2.507 & 5.495 \\
\bottomrule
\end{tabular}
\end{table}

\rev{Table~\ref{tab:cavity_kn_highmom_errors} reports global relative $L_2$ errors and RMSE values for normalized high-order fields. For the normalized fields, the RMSE column is expressed as $100\sqrt{\langle(\mathrm{ML}-\mathrm{Physics})^2\rangle}$, which is more stable than a pointwise relative error in regions where the physical diagnostic approaches zero. The errors increase from $Kn=0.5$ to $Kn=1.0$, as expected for conditions that are extrapolative in the nominal flow parameters. The good performance is therefore interpreted as overlap in the local distribution manifold rather than as unrestricted parameter-space transfer; the larger DM6 error marks the onset of weaker overlap.}

\subsubsection{Component-level profiling}
\rev{Table~\ref{tab:cavity_component_profile} reports a component-level timing breakdown for the q-weighted cavity run. The deterministic physics baseline spends a large fraction of time in full high-order moment evaluation, while the ML mode replaces this by a much cheaper low-order moment calculation plus a small DNN forward pass. The dense $9\times9$ solve is not, by itself, the dominant cost; the main removed cost is the full high-order particle-moment gathering needed to assemble the deterministic closure.}
\rev{The measured cavity speedup is therefore lower than a naive Amdahl estimate based only on deleting the exact-closure block: the ML mode must still perform the lite low-order moment pass and the GPU-native DNN forward pass, which together refill part of the budget vacated by high-order moment assembly and the dense solve. In Table~\ref{tab:cavity_component_profile}, these replacement costs account for 16.80 s and 7.79 s, respectively, while velocity evolution and particle/boundary work remain essentially unchanged. This accounting explains why the realized online speedup in Table~\ref{tab:qweighted_break_even} is configuration-dependent rather than a universal closure-removal bound.}

\begin{table}[t]
\centering
\caption{Component-level runtime profile of the q-weighted cavity simulation. Timings are estimated from synchronized sampled calls for a 50$\times$50 grid, 2.5 million particles, and 3000 time steps. The ``full high-order moments'' entry includes the low-order moment pass plus the extra high-order moment gathering; comparing it with the ML low-order pass gives an implied additional high-order gathering cost of approximately 19.7 s.}
\label{tab:cavity_component_profile}
\setlength{\tabcolsep}{5pt}
\renewcommand{\arraystretch}{1.12}
\begin{tabular}{llcc}
\toprule
Run & Component & Estimated time (s) & Phase fraction (\%) \\
\midrule
Physics & Boundary treatment & 4.82 & 4.8 \\
Physics & Build $9\times9$ system & 5.59 & 5.6 \\
Physics & Full high-order moments & 36.53 & 36.6 \\
Physics & Solve $9\times9$ system & 0.56 & 0.6 \\
Physics & Velocity evolution & 49.90 & 50.0 \\
Physics & Particle move + overhead & 1.33 & 1.3 \\
\midrule
ML & Boundary treatment & 4.82 & 6.0 \\
ML & DNN forward pass & 7.79 & 9.6 \\
ML & Lite low-order moments & 16.80 & 20.7 \\
ML & Velocity evolution & 49.69 & 61.4 \\
ML & Particle move + overhead & 0.86 & 1.1 \\
\bottomrule
\end{tabular}
\end{table}

\subsubsection{Conservation, stability, entropy proxy, and noise sensitivity}
\rev{The DNN surrogate predicts the closure coefficients but does not explicitly impose a discrete H-theorem or a realizability projection. We therefore report direct diagnostics rather than asserting a formal entropy theorem. Table~\ref{tab:qweighted_stability} shows that the ML and physics runs preserve mass to machine precision over the sampled interval, produce no non-finite particles, fields, or coefficients, and do not produce non-positive density, temperature, or second-order moment cells. The collision update itself retains the conservative structure of the cubic-FP particle step: mass is unchanged by construction, momentum is recentered through the local mean velocity, and the ensemble kinetic energy is controlled by the same $\alpha$ rescaling used in the deterministic solver. The predicted closure coefficients enter through the stochastic velocity increment; thus, in the cavity problem the remaining momentum and energy changes are dominated by the moving-wall boundary fluxes rather than by a separate source term from the neural inference. No coefficient clipping or projection was required in the reported q-weighted cavity runs. Table~\ref{tab:cavity_entropy_proxy} reports a binned relative-entropy proxy. This diagnostic is not a proof of an H-theorem; it is a numerical check that the ML run tracks the physics baseline without producing non-finite entropy cells or a systematic excess entropy defect.}

\begin{table}[t]
\centering
\caption{Conservation and stability diagnostics for the q-weighted cavity simulation. For the moving-lid cavity, momentum and energy are affected by wall exchange; therefore mass drift, non-finite counts, positivity, and coefficient boundedness are the primary stability diagnostics.}
\label{tab:qweighted_stability}
\begin{tabular}{lcc}
\hline
Diagnostic & Physics & ML \\
\hline
Max. $|\Delta M/M_0|$ & 0.000e+00 & 0.000e+00 \\
Final $\Delta M/M_0$ & 0.000e+00 & 0.000e+00 \\
Max. non-finite count & 0 & 0 \\
Max. non-positive $\rho,T,\mathrm{DM2}$ cells & 0 & 0 \\
Max. non-finite coefficient count & 0 & 0 \\
Max. $|C|$ & 2.214e+06 & 2.130e+06 \\
Max. $|\Gamma|$ & 4.960e+03 & 4.882e+03 \\
\hline
\end{tabular}
\end{table}

\rev{The coefficient magnitudes in Table~\ref{tab:qweighted_stability} are reported as diagnostics in the nondimensional cavity solver units and should not be read as a universal admissibility or clipping threshold. The large maximum $|C|$ values occur in localized high-gradient or low-support cells but remain finite and did not generate non-positive density, temperature, or second-moment cells in the reported rollout. The $10^6$ coefficient filter used during cylinder data construction is a separate data-quality filter for ill-conditioned cylinder training samples; it is not a global realizability bound applied to all flows or to the cavity diagnostics.}

\begin{table}[t]
\centering
\caption{Discrete entropy-proxy diagnostics for the q-weighted cavity case. The diagnostic is a binned relative-entropy proxy and is not a proof of an H-theorem.}
\label{tab:cavity_entropy_proxy}
\setlength{\tabcolsep}{6pt}
\renewcommand{\arraystretch}{1.12}
\begin{tabular}{lccc}
\toprule
Run & Initial global proxy & Final global proxy & Max. non-finite cells \\
\midrule
Physics & $2.2446\times10^{-1}$ & $2.2419\times10^{-1}$ & 0 \\
ML & $2.2384\times10^{-1}$ & $2.2462\times10^{-1}$ & 0 \\
ML--Physics & -- & $4.2928\times10^{-4}$ & -- \\
\bottomrule
\end{tabular}
\end{table}

\rev{Because particle-based moment estimates contain statistical noise, the study also includes a particle-per-cell sensitivity test. Table~\ref{tab:qweighted_sensitivity} shows that the surrogate remains stable as the particle count is varied, while the expected reduction in noise is observed for most macroscopic quantities. The residual high-order differences are concentrated in the strongest nonequilibrium regions, consistent with the contour comparisons in Figs.~\ref{fig:cavity_highmom_kn1} and~\ref{fig:cavity_highmom_kn05}.}

\begin{table}[t]
\centering
\caption{PPC sensitivity of the q-weighted cavity surrogate for seed 101. This table isolates particle-per-cell effects; seed-to-seed variability is assessed separately in the split-replica and entropy/noise diagnostics.}
\label{tab:qweighted_sensitivity}
\begin{tabular}{cccccc}
\hline
PPC & Seed & $T$ err. (\%) & $|\mathbf{U}|$/$\mathbf{U}$ err. (\%) & $\mathbf{q}$ err. (\%) & $\|\mathbf{q}\|$ err. (\%) \\
\hline
500 & 101 & 0.274 & 1.399 & 1.691 & 3.319 \\
1000 & 101 & 0.203 & 0.990 & 1.190 & 2.428 \\
1500 & 101 & 0.183 & 0.885 & 1.084 & 2.717 \\
\hline
\end{tabular}
\end{table}

\subsubsection{Direct time-averaged closure-coefficient audit}
\rev{To directly quantify the learned closure map, we performed a coefficient audit for an in-manifold cavity condition at $Kn=0.15$ and $U_{\rm lid}=400$ m/s. At sampled post-transient steps, the cubic-FP coefficients were obtained from the full high-order moment system and the local $9\times9$ solve, while the DNN coefficients were evaluated from the same cell state using the GPU-native forward pass. This diagnostic compares the closure coefficients themselves rather than only the evolved macroscopic fields.}

\begin{figure}[H]
    \centering
    \includegraphics[width=0.82\textwidth]{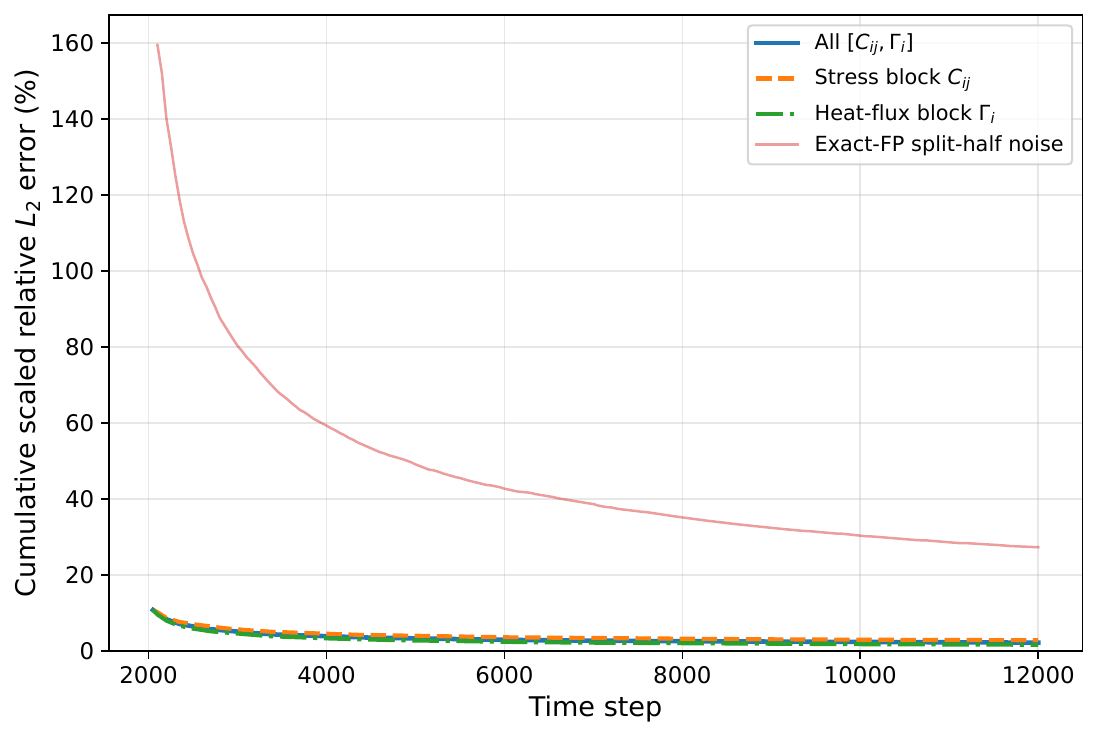}
    \caption{Time history of the cumulative time-averaged closure-coefficient error for the in-manifold cavity audit at $Kn=0.15$ and $U_{\rm lid}=400$ m/s. The error is reported in the standardized output space used during DNN training.}
    \label{fig:cavity_closure_coeff_timeavg_history}
\end{figure}

\rev{Because the instantaneous cubic-FP coefficients are computed from high-order particle moments, they contain particle-noise fluctuations. We therefore report the cumulative time-averaged coefficient error in the standardized output space used during DNN training. Table~\ref{tab:cavity_closure_coeff_timeavg} shows that the final cumulative errors are below $5\%$ for the full coefficient vector, the stress-relaxation block $C_{ij}$, and the heat-flux block $\Gamma_i$. Figure~\ref{fig:cavity_closure_coeff_timeavg_history} shows that the cumulative error decreases during the post-transient averaging window and remains bounded. This audit confirms that, on the representative training manifold, the GPU-native surrogate reproduces the time-averaged cubic-FP closure coefficients with controlled error.}

\rev{The direct coefficient-level audit is summarized in Table~\ref{tab:cavity_closure_coeff_timeavg}. The table reports the mean and final cumulative standardized $L_2$ errors for the complete closure vector, the stress-relaxation block $C_{ij}$, and the heat-flux block $\Gamma_i$, confirming that the learned closure remains within a few percent of the exact cubic-FP coefficient solve for the in-manifold cavity case.}

\begin{table}[t]
\centering
\caption{Direct time-averaged closure-coefficient audit for the in-manifold cavity case at $Kn=0.15$ and $U_{\rm lid}=400$ m/s. Errors are computed in the standardized output space used during DNN training.}
\label{tab:cavity_closure_coeff_timeavg}
\small
\setlength{\tabcolsep}{6pt}
\renewcommand{\arraystretch}{1.15}
\begin{tabular}{lcc}
\toprule
Coefficient block & Mean cumulative $\varepsilon_{L_2}^{(s)}$ (\%) & Final cumulative $\varepsilon_{L_2}^{(s)}$ (\%) \\
\midrule
All coefficients $[C_{ij},\Gamma_i]$ & 3.319 & 2.248 \\
Stress-relaxation block $C_{ij}$ & 3.958 & 2.892 \\
Heat-flux block $\Gamma_i$ & 2.791 & 1.698 \\
\bottomrule
\end{tabular}
\end{table}

\subsubsection{Online speedup and break-even analysis}
\rev{The speedup is configuration-dependent because the attainable acceleration depends on the grid, particle count, and the remaining non-closure costs. Table~\ref{tab:qweighted_break_even} also includes the offline data-generation and training costs. For the measured q-weighted cavity setup, the online saving is about 121 s per run. The training-only cost is recovered after roughly 10 target simulations, while the full offline cost including data generation is recovered after roughly 26 target simulations. This analysis clarifies that the method is most useful for parameter studies or repeated simulations after the offline cost has been amortized.}

\begin{table}[t]
\centering
\caption{Break-even analysis for the q-weighted GPU-native surrogate. The online saving is measured from the cavity run, while the offline cost includes data generation and network training when available.}
\label{tab:qweighted_break_even}
\begin{tabular}{lc}
\hline
Quantity & Value \\
\hline
Physics online time per run & 664.65 s \\
ML online time per run & 543.29 s \\
Online saving per run & 121.36 s \\
Online speedup & 1.223$\times$ \\
Training cost & 1155.61 s \\
Data-generation cost & 2016.60 s \\
Full offline cost & 3172.21 s \\
Break-even, training only & 9.52 runs \\
Break-even, full offline cost & 26.14 runs \\
\hline
\end{tabular}
\end{table}

\section{Extension to External Aerodynamics: Hypersonic Flow over a Cylinder}

\rev{To evaluate stability and closed-loop accuracy in a shock-dominated external flow, we extend the validation domain from internal flows to high-speed rarefied flow over a cylinder. This case is not used as a cross-condition generalization test; it is a same-condition closed-loop rollout of the deployed surrogate in a flow with a detached bow shock, a stagnation point with strong thermodynamic gradients, a curvilinear solid boundary, and a rarefied wake region. These features provide a stringent stress test of whether the learned closure remains stable and tracks the exact cubic-FP reference under the tested cylinder condition.}

\subsection{Problem Description and Physics-Based Data Generation}
The simulation setup considers argon flow over a cylinder of diameter $D = 0.3048$ m, a widely used rarefied hypersonic benchmark~\cite{lofthouse2008velocity,goshayeshi2015dsmc,goshayeshi2015novel,jun2018assessment}. The free-stream conditions are $U_\infty = 2624$ m/s and $T_\infty = 200$ K. For monatomic argon ($\gamma=5/3$), these values correspond to $M_\infty \simeq 10.0$. The freestream number density used in the reference setup is $n_\infty = 4.247\times10^{20}\,\mathrm{m}^{-3}$, corresponding to $Kn_\infty \simeq 0.01$ based on the cylinder diameter. The cylinder wall temperature is isothermal at $T_{wall}=500$ K. The computational domain extends radially to $1.5D$, providing clearance to capture the detached bow shock and immediate wake.

To generate a high-fidelity training dataset, we employed the baseline cubic Fokker-Planck (FP) solver. To efficiently capture the multi-scale physics of the high-speed rarefied flow, particularly the detached bow shock and the steep gradients within the boundary layer, a dynamic Adaptive Mesh Refinement (AMR) strategy is employed. The spatial domain is discretized using a hierarchical Cartesian Quadtree structure. The grid adaptation is driven by both local particle density and macroscopic flow gradients, ensuring high resolution in regions of interest while minimizing computational cost in the far field.

The refinement decision for a computational cell $i$ is governed by a composite criterion based on the normalized gradients of density ($\rho$), temperature ($T$), and thermodynamic pressure ($P$). Let $\phi \in \{\rho, T, P\}$ represent a macroscopic flow variable. The normalized deviation, $\Delta \phi_i$, between the cell $i$ and its parent node is defined as:

\begin{equation}
    \Delta \phi_i = \frac{|\phi_i - \phi_{\text{parent}}|}{|\phi_i| + \epsilon},
\end{equation}

\noindent where $\epsilon$ is a small regularization constant to prevent division by zero. A cell is flagged for refinement (splitting into four children) if any of the following conditions are met:

1. Gradient Constraint: The maximum normalized deviation exceeds a user-defined threshold, $\tau_{\text{grad}}$ (set to $0.10$ in this study):
    \begin{equation}
        \max(\Delta \rho_i, \Delta T_i, \Delta P_i) > \tau_{\text{grad}}.
    \end{equation}
    
2. Particle Density Constraint: The number of computational particles within the cell, $N_{p,i}$, exceeds a maximum limit, $N_{\text{max}}$, to strictly control the statistical noise and prevent load imbalance.

3. Shock Detection: To explicitly capture the strong bow shock, cells are refined if the local pressure exceeds the free-stream pressure by a specific factor (e.g., $P_i > 1.2 P_{\infty}$), regardless of local gradients.

Conversely, a coarsening (merging) operation is performed to reduce resolution in the wake or far-field regions where flow variations are minimal. A block of four sibling cells is merged back into their parent if the normalized gradients among them fall below a coarsening threshold, $\tau_{\text{coarse}}$ (typically $0.05$), and the total particle count is sufficiently low. To prevent numerical oscillation (flickering) of the grid, a hysteresis mechanism based on "cell age" is implemented; cells are only permitted to coarsen if they have remained static for a predefined number of time steps (e.g., 2000 steps).

We extracted approximately 5 million sampling points from the baseline simulation to cover the bow shock, stagnation layer, near-wall gas, and wake. Samples with non-finite values or extremely large closure coefficients ($|C_{ij}|$ or $|\Gamma_i|>10^6$ in the solver's nondimensional coefficient units) were removed as a data-quality filter for ill-conditioned, low-count outliers. This filtering is applied only during dataset construction; the deployed ML solver does not clip or project the network output. The cylinder test is therefore a closed-loop rollout and self-consistency test at the same freestream condition used to construct the cylinder training data, not a demonstration of cross-condition cylinder generalization. This distinction is important: the cavity section tests held-out complete conditions, whereas the cylinder section tests whether the learned closure remains stable and accurate when inserted back into the coupled shock-layer particle solver for thousands of time steps.

The cylinder surrogate is a fully connected MLP with 16 standardized inputs and 9 outputs. The outputs are the six independent components of $C_{ij}$ and the three components of $\Gamma_i$; no additional geometry-dependent closure coefficient is introduced. The 16 inputs are low-order local moments and solver variables available online: density, temperature, velocity, stress, heat flux, pressure, and relaxation-time information. These inputs do not uniquely specify an arbitrary kinetic distribution, so the learned closure is interpreted as a regression on the training manifold rather than as a universal kinetic map.

For cylinder deployment, the coefficients are coupled into the time-step with a temporal under-relaxation factor $w=0.2$. This is a numerical coupling parameter used to damp step-to-step coefficient noise in the shock layer; it is not a realizability projection and it does not enforce positivity. The coefficient field used in the next stochastic update is the weighted average of the new network prediction and the previous coefficient field. This smoothing is an important part of the closed-loop cylinder deployment, and all cylinder flow-field, surface-coefficient, entropy-audit, and timing results should be interpreted as results for this deployed $w=0.2$ configuration. It is disclosed separately because the q-weighted cavity runs reported above used raw network outputs without clipping or projection. We do not interpret the cylinder rollout as evidence that closed-loop stability is independent of this numerical coupling choice.

\subsection{Results and Validation}

\begin{figure}[!htbp]
    \centering
    \includegraphics[width=0.7\linewidth]{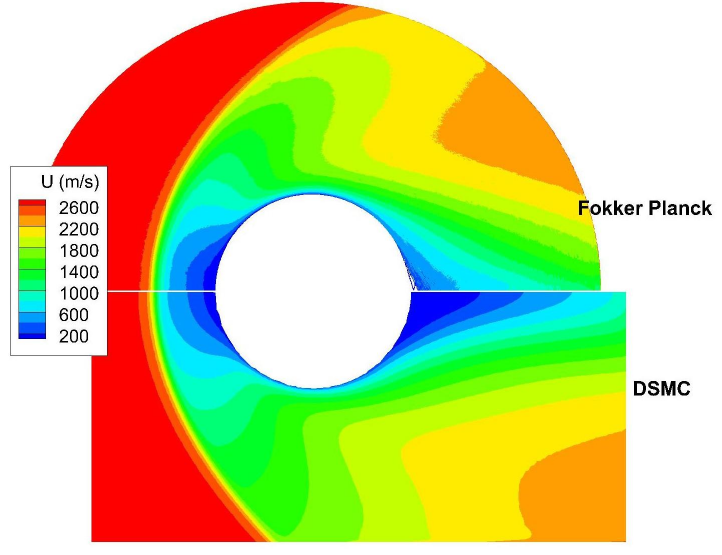}
    
    \caption{Validation of the Fokker-Planck model against DSMC benchmark data for hypersonic flow over a cylinder.}
    \label{fig:fp_dsmc_validation}
\end{figure}

First, we validate our FP solution for the hypersonic cylinder flow with the DSMC solution of the same test case~\cite{goshayeshi2015dsmc}. The contours shown in Fig.~\ref{fig:fp_dsmc_validation} represent the velocity magnitude ($U$). The top half displays the results from the present Fokker-Planck solver, while the bottom half shows the DSMC solution, demonstrating strong agreement in capturing the bow shock structure and the wake region.

The predictive fidelity of the ML surrogate was evaluated by comparing its steady-state solution against the full physics-based solver. The results in the following sections indicate that the surrogate captures the main macroscopic features of the rarefied hypersonic flow.

\subsubsection{Flow Field Contours}
\begin{figure}[!htbp]
    \centering
    \includegraphics[width=0.7\textwidth]{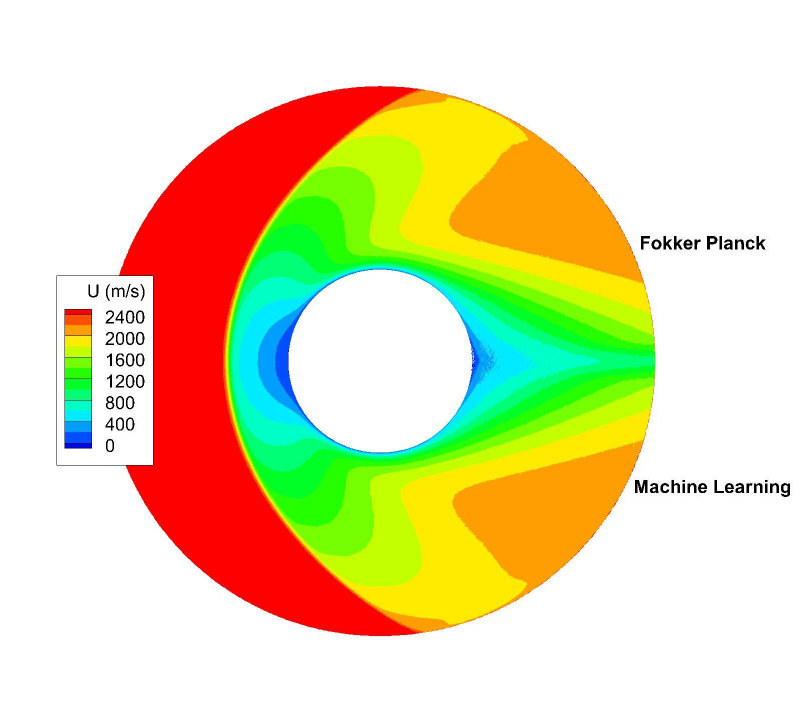}
    \caption{Comparison of the velocity magnitude field (m/s).}
    \label{fig:U_contours}
\end{figure}

\begin{figure}[!htbp]
    \centering
    \includegraphics[width=0.7\textwidth]{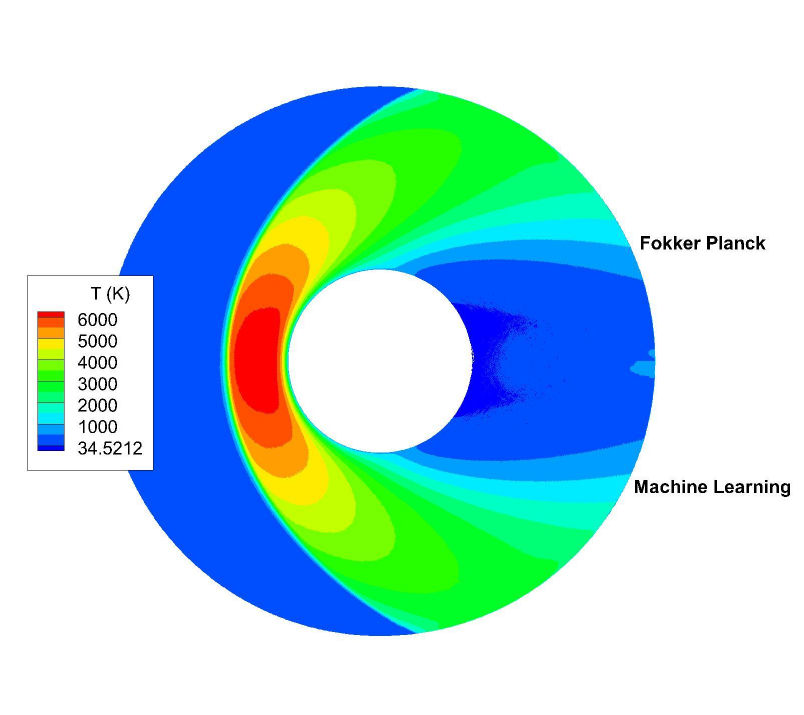}
    \caption{Comparison of the temperature field (K) between the Fokker-Planck physics solver (top half) and the Machine Learning surrogate (bottom half).}
    \label{fig:T_contours}
\end{figure}

The macroscopic flow structure is visualized in the contour plots of velocity and temperature magnitude. 
Figure \ref{fig:U_contours} illustrates the velocity magnitude contours, where the top half represents the physics-based solution and the bottom half displays the ML surrogate prediction. The surrogate captures the detached bow-shock structure, including its standoff distance and curvature, for the cylinder condition represented in the training data. This is a stringent closed-loop test because the shock position is sensitive to stress and heat-flux relaxation. The comparison should be interpreted as evidence of accuracy on the validated cylinder regime, not as proof of unrestricted transferability to arbitrary hypersonic or high-Knudsen conditions.
Figure \ref{fig:T_contours} presents the temperature field comparison. The high-temperature region in the stagnation zone, resulting from the conversion of kinetic energy into internal energy, is also reproduced consistently with the exact cubic-FP reference for the tested rollout.

\subsubsection{Surface Aerodynamic Properties}
For practical engineering applications, the accurate prediction of surface loads and heat transfer is important. We extracted the pressure coefficient ($C_p$), skin friction coefficient ($C_f$), and heat transfer coefficient ($C_h$) along the cylinder surface to assess the consistency of surface loads and heat transfer for the same-condition cylinder rollout.

\begin{figure}[!htbp]
    \centering
    \begin{subfigure}[b]{0.48\textwidth}
        \includegraphics[width=\textwidth]{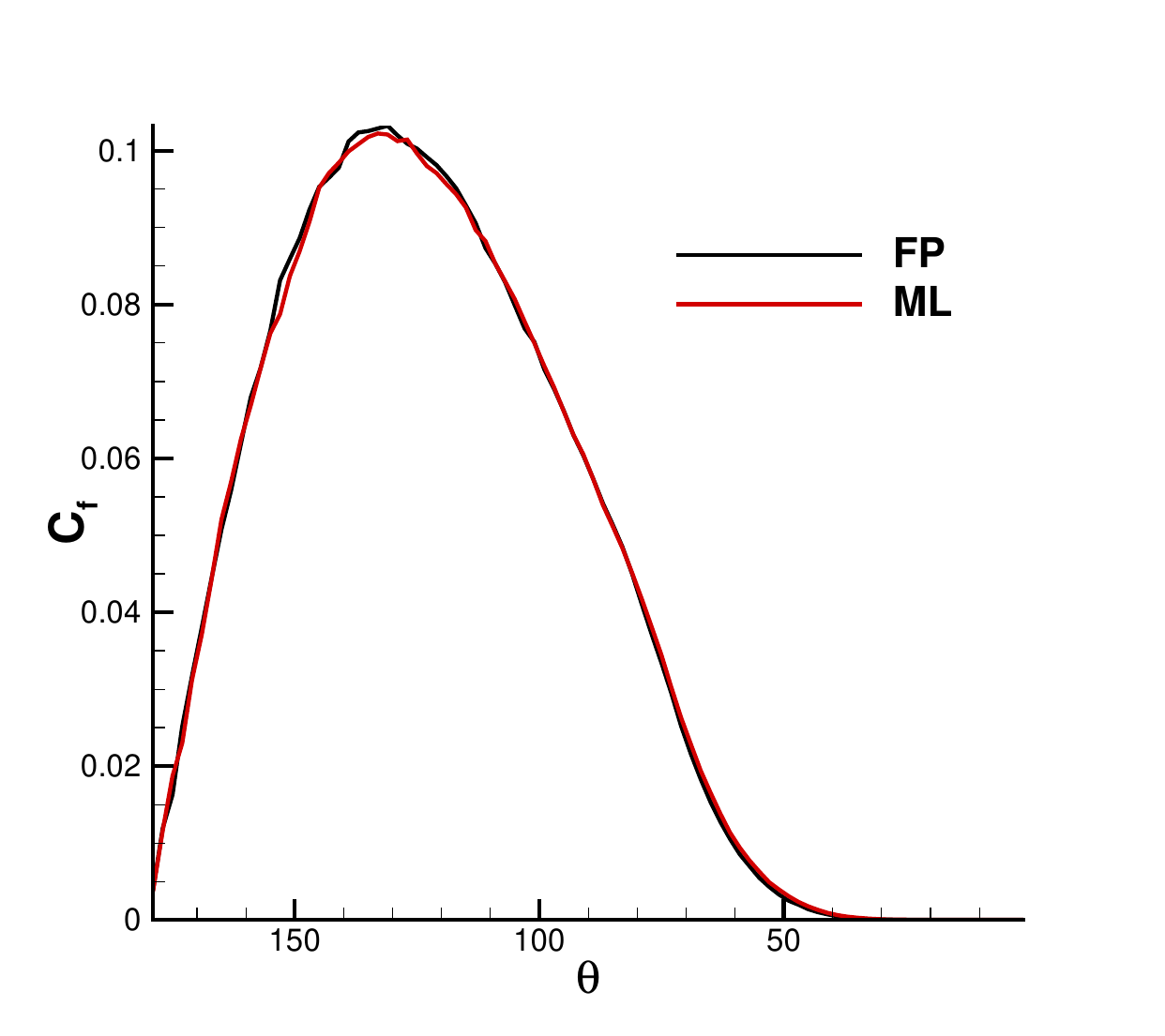}
        \caption{Skin Friction Coefficient ($C_f$)}
        \label{fig:Cf}
    \end{subfigure}
    \hfill
    \begin{subfigure}[b]{0.48\textwidth}
        \includegraphics[width=\textwidth]{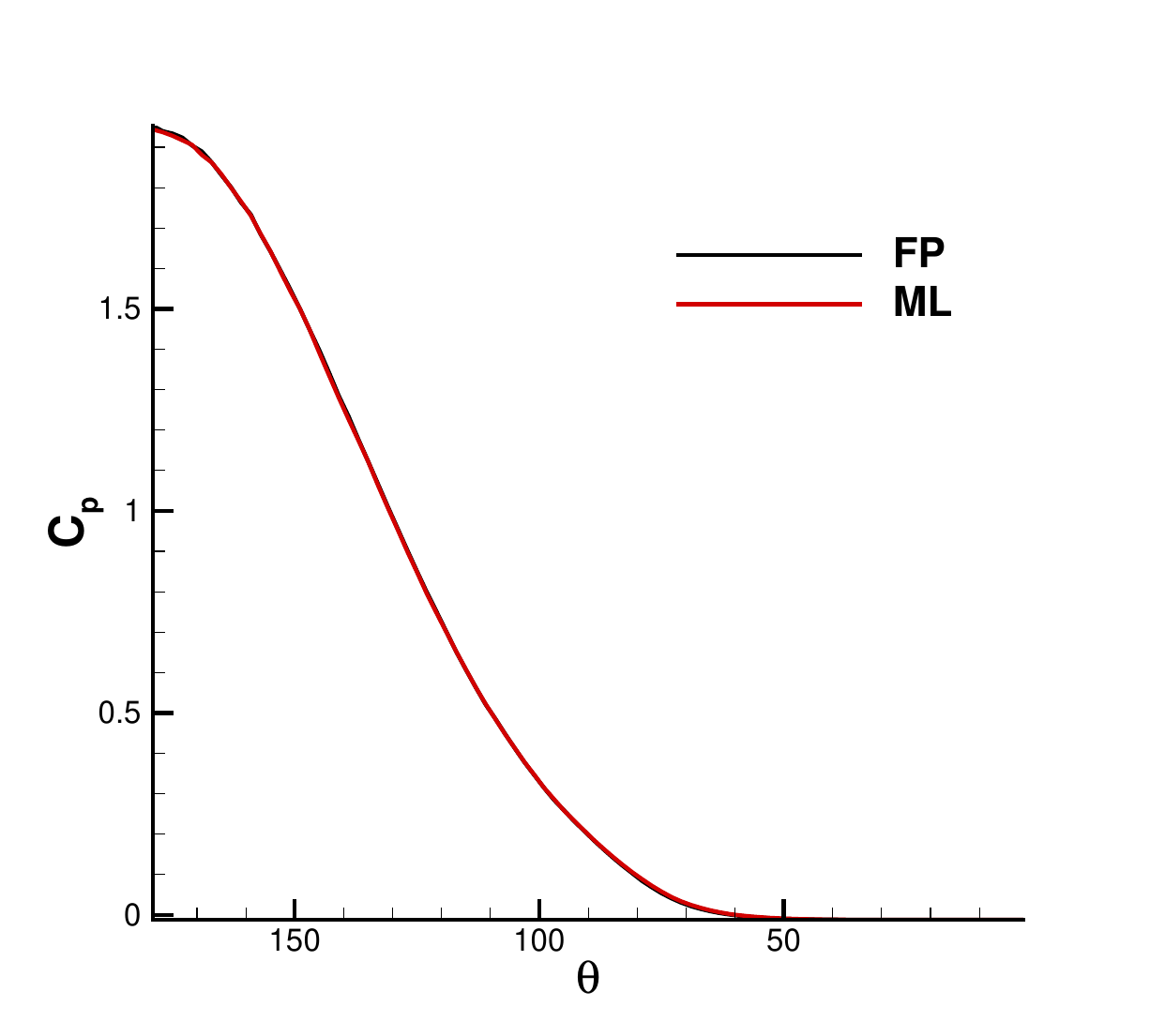}
        \caption{Pressure Coefficient ($C_p$)}
        \label{fig:Cp}
    \end{subfigure}
    \caption{Surface aerodynamic coefficients along the cylinder wall for the same-condition cylinder rollout. The red dashed lines (ML) closely track the black solid lines (Physics) for the tested benchmark; $\theta$ follows the aerodynamic-surface convention with the front stagnation point at $180^\circ$.}
    \label{fig:Cp_Cf}
\end{figure}

\begin{figure}[!htbp]
    \centering
    \includegraphics[width=0.6\textwidth]{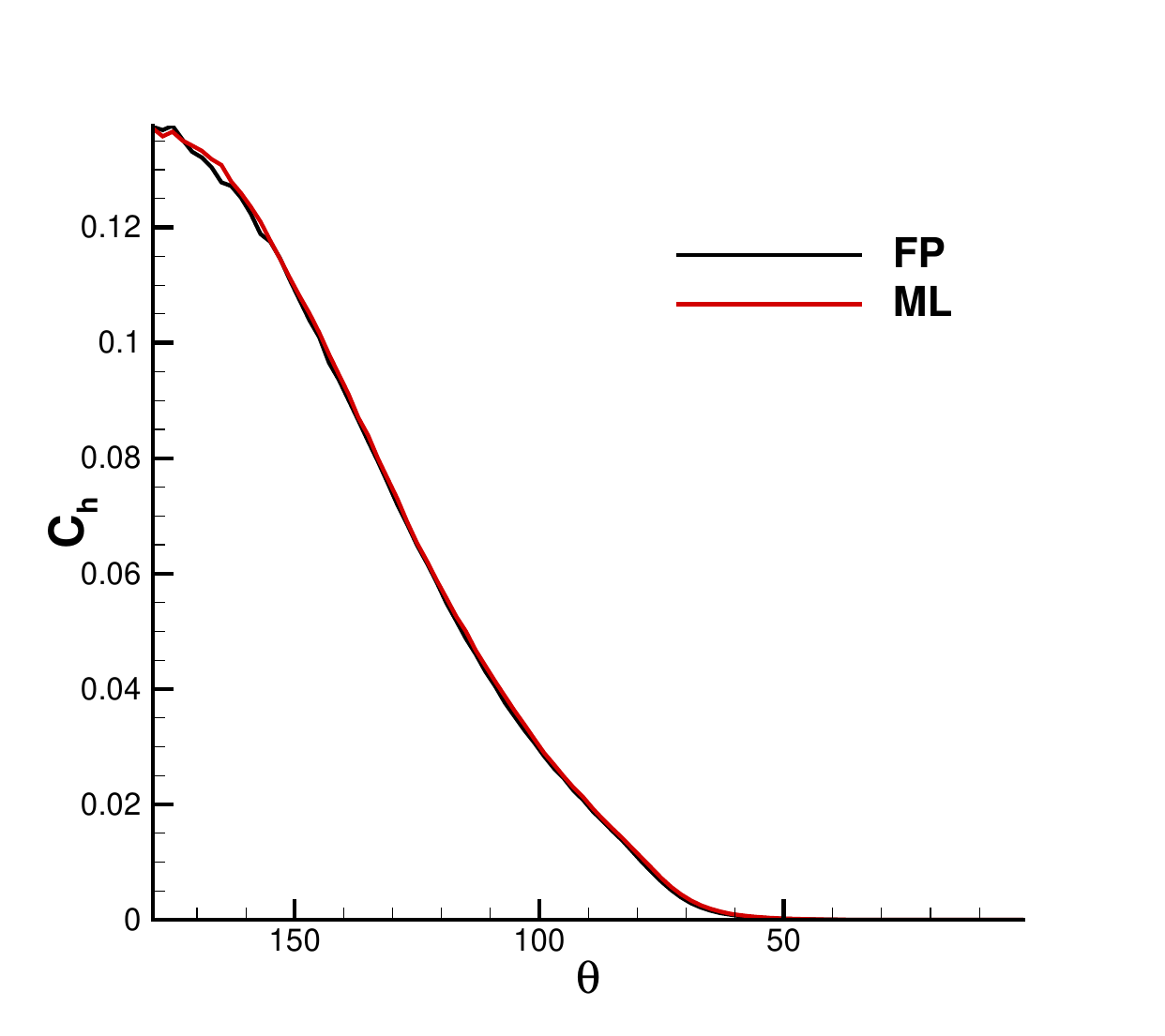}
    \caption{Heat-transfer coefficient ($C_h$) distribution for the same-condition cylinder rollout. The surrogate follows the exact cubic-FP stagnation-heating peak and decay trend for the tested condition.}
    \label{fig:Ch}
\end{figure}

\rev{Figure~\ref{fig:Cp_Cf} displays the skin-friction and pressure coefficients. The aerodynamic surface angle $\theta$ in Figs.~\ref{fig:Cp_Cf}--\ref{fig:Ch} follows the conventional cylinder-wall convention used in the original data, with the front stagnation point at $\theta=180^\circ$. The pressure coefficient profile exhibits the expected hypersonic trend, peaking at the stagnation point and decreasing as the flow accelerates around the cylinder. The ML predictions closely overlap the physics baseline for this same-condition cylinder rollout, indicating that the deployed closure preserves the surface-load trends for the tested benchmark.}

\rev{Furthermore, Figure~\ref{fig:Ch} presents the heat-transfer coefficient distribution. Predicting aerodynamic heating is difficult in rarefied flows because stress and heat-flux relaxation are coupled. The ML curve reproduces the stagnation-region peak and subsequent decay trend of the exact cubic-FP reference for the tested rollout, while the quantitative validation burden is carried by the entropy and coefficient-profile audits below.}

\rev{Table~\ref{tab:cylinder_surface_coeff_check} summarizes the role of the three surface-coefficient comparisons. These curves are used as a qualitative engineering-load check for the same-condition cylinder rollout, while the entropy-profile and coefficient-audit diagnostics below provide the main quantitative consistency evidence for the cylinder surrogate.}

\begin{table}[t]
\centering
\caption{Cylinder surface-coefficient comparison used as a qualitative engineering-load check. The entries summarize the plotted exact-FP/ML-FP curve agreement; the entropy-profile and coefficient-audit tables provide the main quantitative consistency tests for the cylinder surrogate.}
\label{tab:cylinder_surface_coeff_check}
\small
\renewcommand{\arraystretch}{1.15}
\begin{tabularx}{\textwidth}{@{}l >{\raggedright\arraybackslash}X >{\raggedright\arraybackslash}X@{}}
\toprule
Quantity & Observed comparison & Role in validation \\
\midrule
$C_p$ & ML curve overlaps the exact-FP stagnation peak and shoulder-decay trend. & Surface-load consistency \\
$C_f$ & ML curve follows the exact-FP wall-shear trend. & Near-wall shear check \\
$C_h$ & ML curve follows the heating peak and downstream-decay trend. & Thermal-load consistency \\
\bottomrule
\end{tabularx}
\end{table}

\subsubsection{Covariance-based entropy-proxy audit on the stagnation line and cylinder surface}
\label{sec:cylinder_entropy_audit}
\reventropy{Entropy-related behavior and stability are especially important for the hypersonic cylinder because the detached bow shock and stagnation layer are the strongest nonequilibrium regions in this paper. We therefore include a dedicated covariance-based entropy-proxy audit for this case. We report entropy levels along selected profiles rather than local entropy-production rates, because production-rate estimates from particle data require noisy temporal/spatial derivatives and are not robust at the present bin resolution. This audit is not a separate neural-network model: the deployed network is the same GPU-native $C/\Gamma$ closure surrogate used in the cylinder flow-field comparisons. No entropy quantity is fitted or predicted by the DNN. Instead, the exact-FP and ML-FP solvers are rolled out independently, and the entropy diagnostics are computed a posteriori from the resulting particle distributions and macroscopic density fields. This distinction tests the closed-loop effect of the learned closure on the evolved kinetic state rather than only testing a post-processed coefficient fit.}

\reventropy{The starting point is the Boltzmann entropy of a dilute gas, $H=\int f\log f\,d\bm v$, or equivalently the entropy density $-k_B\int f\log f\,d\bm v$ up to an additive constant \cite{Bird1994,cercignani2000rarefied}. The full velocity distribution $f(\bm x,\bm v)$ is not stored as a resolved histogram in the production GPU solver, so we use the standard moment-consistent decomposition into a local equilibrium part and a non-equilibrium anisotropy deficit. Let $\bm c=\bm v-\bm U$ be the peculiar velocity and let}
\begin{equation}
\Theta_{ij}=\langle c_i c_j\rangle, \qquad
\theta=\frac{1}{3}{\rm tr}(\Theta),
\label{eq:cyl_theta_def}
\end{equation}
\reventropy{where the brackets denote a particle average in a profile bin. The scalar $\theta$ is proportional to the translational temperature. For a fixed monatomic species, the entropy per particle of a Maxwellian with density $\rho$ and scalar temperature $\theta$, nondimensionalized by $k_B$, contains the Sackur--Tetrode/ideal-gas dependence $\frac{3}{2}\log\theta-\log\rho$ plus constants that cancel when profiles are normalized by the upstream state. Natural logarithms are used throughout. 
The following entropy-proxy construction uses standard kinetic-theory and information-theoretic identities: the local-equilibrium term follows from the ideal-gas/Sackur--Tetrode entropy difference for a monatomic gas \cite{Bird1994,cercignani2000rarefied}, while the anisotropy deficit follows from the closed-form Kullback--Leibler divergence between Gaussian distributions with covariance tensors $\Theta$ and $\theta I$. The covariance-based maximum-entropy interpretation is consistent with classical moment-closure ideas \cite{Grad1949,Levermore1996}. The contribution of the present work is to use these standard identities as a covariance-based exact-FP/ML-FP fidelity audit for the evolved cylinder particle distributions.
We therefore compute}
\begin{equation}
\frac{\Delta s_{\rm eq}}{k_B}
= \frac{3}{2}\log\left(\frac{\theta}{\theta_\infty}\right)
-\log\left(\frac{\rho}{\rho_\infty}\right),
\label{eq:cyl_entropy_eq}
\end{equation}
\reventropy{where $(\theta_\infty,\rho_\infty)$ are measured from the upstream front-line reference bins. This is the entropy of the local Maxwellian having the same density and scalar temperature as the sampled state. It is not by itself a non-equilibrium entropy correction; it isolates the thermodynamic contribution associated with compression and heating.}

\reventropy{This surface quantity should not be interpreted in the same way as the pressure or heat-transfer coefficients, which peak at the front stagnation point. Equation~\eqref{eq:cyl_entropy_eq} is an upstream-normalized near-wall gas entropy proxy containing both the heating contribution $\frac{3}{2}\log(\theta/\theta_\infty)$ and the compression contribution $-\log(\rho/\rho_\infty)$. In the stagnation layer, strong compression can offset part of the temperature increase in this per-particle entropy measure, whereas downstream near-wall gas can be less compressed while still thermally disturbed by the shock layer and wall interaction. The resulting surface trend is therefore a local gas-state entropy-proxy audit, not a stagnation-peak surface-load quantity, wall entropy, or entropy-production profile.}

\reventropy{To quantify the departure from a scalar-temperature Maxwellian without constructing a noisy velocity histogram, we use the Gaussian entropy deficit associated with the measured covariance tensor. For a Gaussian distribution with covariance $\Theta$, the Kullback--Leibler divergence to the isotropic Gaussian with the same scalar temperature $\theta I$ is}
\begin{equation}
D_{\rm KL}\!\left({\cal N}(0,\Theta)\,\|\,{\cal N}(0,\theta I)\right)
=\frac{1}{2}\left[{
{\rm tr}}\!\left(\frac{\Theta}{\theta}\right)-3
-\log\det\!\left(\frac{\Theta}{\theta}\right)\right].
\end{equation}
\reventropy{Since ${\rm tr}(\Theta/\theta)=3$ by definition, this reduces to}
\begin{equation}
D_G=-\frac{1}{2}\log\det\left(\frac{\Theta}{\theta}\right)\ge 0,
\label{eq:cyl_DG}
\end{equation}
\reventropy{for positive-definite $\Theta$. This is a moment-based measure of anisotropic non-equilibrium. It is also the Gaussian maximum-entropy correction implied by retaining the second-moment tensor rather than only the scalar temperature; the same covariance-based maximum-entropy logic underlies moment closures for kinetic theory \cite{Grad1949,Levermore1996}. The corresponding Gaussian kinetic-entropy estimate is then}
\begin{equation}
\frac{\Delta s_{{\rm kin},G}}{k_B}
= \frac{\Delta s_{\rm eq}}{k_B} - \left(D_G-D_{G,\infty}\right).
\label{eq:cyl_entropy_kinG}
\end{equation}
\reventropy{The subtraction of $D_{G,\infty}$ removes the small finite-particle upstream bias in the covariance anisotropy estimate. A heat-flux correction $D_q=\|\hat{\bm q}\|^2/5$, with $\hat{\bm q}=\langle \bm c\|\bm c\|^2\rangle/(2\theta^{3/2})$, was also evaluated. Because this term depends on a third-order moment and is much more sensitive to particle noise and tail statistics, it is reported only as a supplemental diagnostic and is not used as the primary entropy claim.}

\reventropy{A separate question is whether the learned closure can be shown to produce a non-negative entropy-production rate. We do not make that claim. The cubic-FP reference used here does not possess a general H-theorem, because the cubic drift is modified while the diffusion remains isotropic; consequently a formal proof of $dS/dt\ge0$ is unavailable for the reference model itself and cannot be recovered by the neural surrogate. Production-rate estimates from particle data would also require noisy temporal and spatial derivatives at the shock and surface bins. We therefore use the strongest reproducible substitute available in this setting: a like-for-like entropy-level fidelity audit. The exact-FP and ML-FP solvers are rolled out independently with identical sampling and binning, and the a posteriori entropy proxies are compared to test whether the neural closure introduces a systematic entropy-proxy defect relative to the exact cubic-FP reference. This is a necessary surrogate-fidelity check, not a second-law proof or a local entropy-production measurement.}

\reventropy{The cylinder audit uses two independent exact-FP replicas and two independent ML-FP replicas in the same particle-resolution regime used for the cylinder surrogate validation ($8.0\times 10^6$ particles, $400\times300$ cells, 4000 steps, 1000 warm-up steps, and sampling every 50 steps). In the ML-FP replica runs, the neural $C/\Gamma$ coefficients are inserted at the same program point where exact coefficients would otherwise be used for broadcast and downstream stochastic updates. This wording refers to the insertion location in the shared solver infrastructure: the coefficient field used for the subsequent $\alpha$/noise construction and particle update is the neural field. Timing claims for the accelerated cylinder path are based on the production ML mode, not on the entropy-audit bookkeeping path. The front profiles are extracted only ahead of the cylinder on the stagnation line. The surface entropy-proxy profiles use a cell-centered near-wall gas layer adjacent to the isothermal cylinder and therefore represent gas entropy in the first sampled layer, not a solid-wall thermodynamic state. The surface display is split into a high-support front-side sector and a support-limited rear-side sector: the front side is retained at the fine angular resolution used in the original gate-reference audit, while the rear/wake side is shown as coarse support-controlled angular bins with explicit uncertainty bars.}

\reventropy{The angular trend of the surface entropy proxy should not be interpreted in the same way as the wall pressure or heat-transfer coefficient. In particular, $\Delta s_{\rm eq}/k_B$ in Eq.~\eqref{eq:cyl_entropy_eq} contains both a scalar-temperature contribution and the density term $-\log(\rho/\rho_\infty)$, and the plotted quantity is evaluated in the first sampled gas layer rather than at the solid wall. As the flow expands around the shoulder and into the rarefied rear-side layer, the decrease in density and the increasing nonequilibrium/anisotropic spread of the sampled gas can increase the covariance-based entropy proxy even though the stagnation point remains the location of maximum aerodynamic pressure and heat-transfer loading. The surface-entropy plots are therefore used only as like-for-like exact-FP/ML-FP fidelity diagnostics under identical binning and normalization, not as a statement that the surface entropy must peak at the stagnation point.}

\reventropy{For reproducibility, the front-line bins have spacing $\Delta(x/R)=0.02$ and use a narrow stagnation-line tube $|y|/R\le0.025$ with Gaussian transverse weighting. The surface entropy proxy is first evaluated on $1^\circ$ near-wall gas samples in the annular layer $0.005\le r/R-1\le0.15$ with radial Gaussian weighting of width $0.05R$; this layer is deliberately a gas sampling layer and is not the wall boundary state. The gas-surface interaction in the underlying solver is diffuse reflection from an isothermal wall at $T_{wall}=500$ K. The covariance quantities $\Theta$, $\theta$, $D_G$, and $D_q$ are computed from particle velocities in each bin, while $\rho$ is computed by cell-volume/geometry weighting of the solver density field to reduce particle-count noise. This mixed estimator is used identically for exact-FP and ML-FP replicas, so it affects only the absolute entropy proxy and not the like-for-like surrogate comparison. Particle weights are included in all bin averages. Bins with non-positive density or temperature, fewer than the valid-sample requirement, or non-finite covariance statistics are masked; the eigenvalues of $\Theta/\theta$ are floored at $10^{-12}$ only when evaluating $\log\det(\Theta/\theta)$ to avoid numerical underflow in nearly singular finite-sample bins. For the rear/wake sector ($0^\circ\le\theta<60^\circ$), the valid $1^\circ$ surface estimates are aggregated into $10^\circ$ coarse angular bins. The error bars plotted in this sector denote the conservative half-range of the valid fine-bin estimates inside each coarse bin after applying the same reference normalization to exact-FP and ML-FP data. They therefore visualize finite-sample, adaptive-binning, and particle-support uncertainty in the rarefied rear/wake region, rather than an additional physical entropy-production estimate.}

\begin{figure}[H]
\centering
\includegraphics[width=0.86\textwidth,height=0.86\textheight,keepaspectratio]{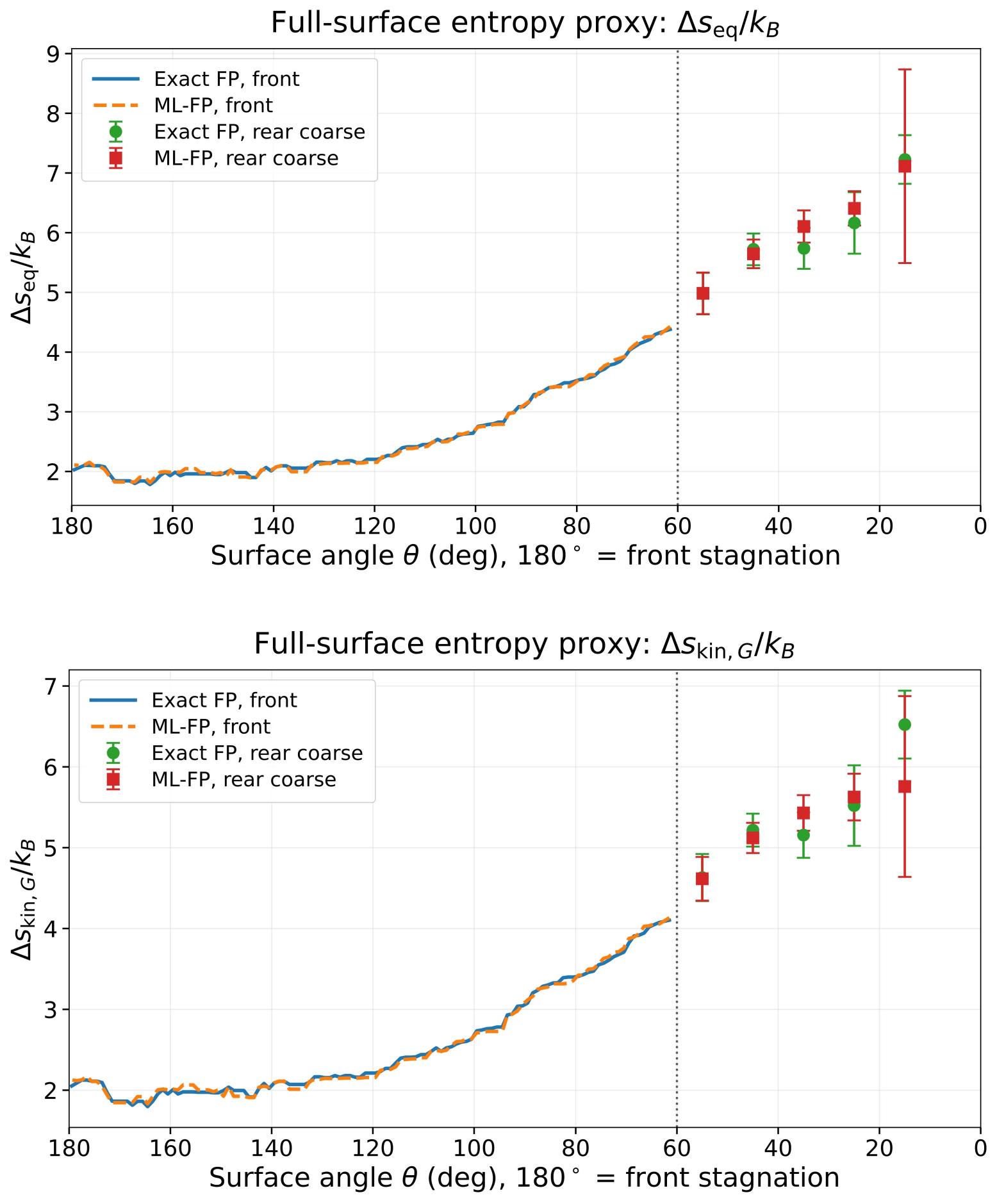}
\caption{\reventropy{Covariance-based entropy-proxy comparison in the near-wall gas layer over the full cylinder surface. The top panel shows $\Delta s_{\rm eq}/k_B$ and the bottom panel shows $\Delta s_{{\rm kin},G}/k_B$. The horizontal axis follows the same aerodynamic surface-angle convention as the cylinder coefficient plots, with $\theta=180^\circ$ at the front stagnation point. For $60^\circ\le\theta\le180^\circ$, the profiles are shown as fine-resolution front-side curves. For the support-limited rear/wake sector $0^\circ\le\theta<60^\circ$, the data are aggregated into coarse angular bins with conservative half-range error bars, which indicate statistical/binning uncertainty caused by sparse particle support, stronger adaptive-mesh/binning sensitivity, and larger temporal-to-spatial sampling variability in the rarefied wake-side sampling layer. The surface profiles are near-wall gas profiles, not wall-state entropy, and they are not interpreted as a monotone surface law or an entropy-production rate.}}
\label{fig:cylinder_entropy_surface}
\end{figure}

\reventropy{The fine-resolution front-side surface bins carry the main quantitative surface-entropy comparison. The rear-side/wake bins are included in Fig.~\ref{fig:cylinder_entropy_surface} for transparency, but only as support-controlled coarse-bin diagnostics. The larger error bars there are expected because the rarefied rear-side near-wall layer has fewer particle hits, stronger adaptive-mesh/binning sensitivity, and larger temporal-to-spatial sampling variability than the front shock/stagnation layer. The Gaussian entropy estimate should also be read as a second-moment maximum-entropy surrogate, not as the full Boltzmann $H$-functional of a possibly bimodal or tail-sensitive wake distribution. The key validation point is therefore the like-for-like exact-FP/ML-FP comparison under identical binning, reference normalization, and particle resolution, not a claim of a monotone surface-entropy law or a rigorous entropy-production rate.}

\begin{figure}[H]
\centering
\includegraphics[width=0.86\textwidth,height=0.86\textheight,keepaspectratio]{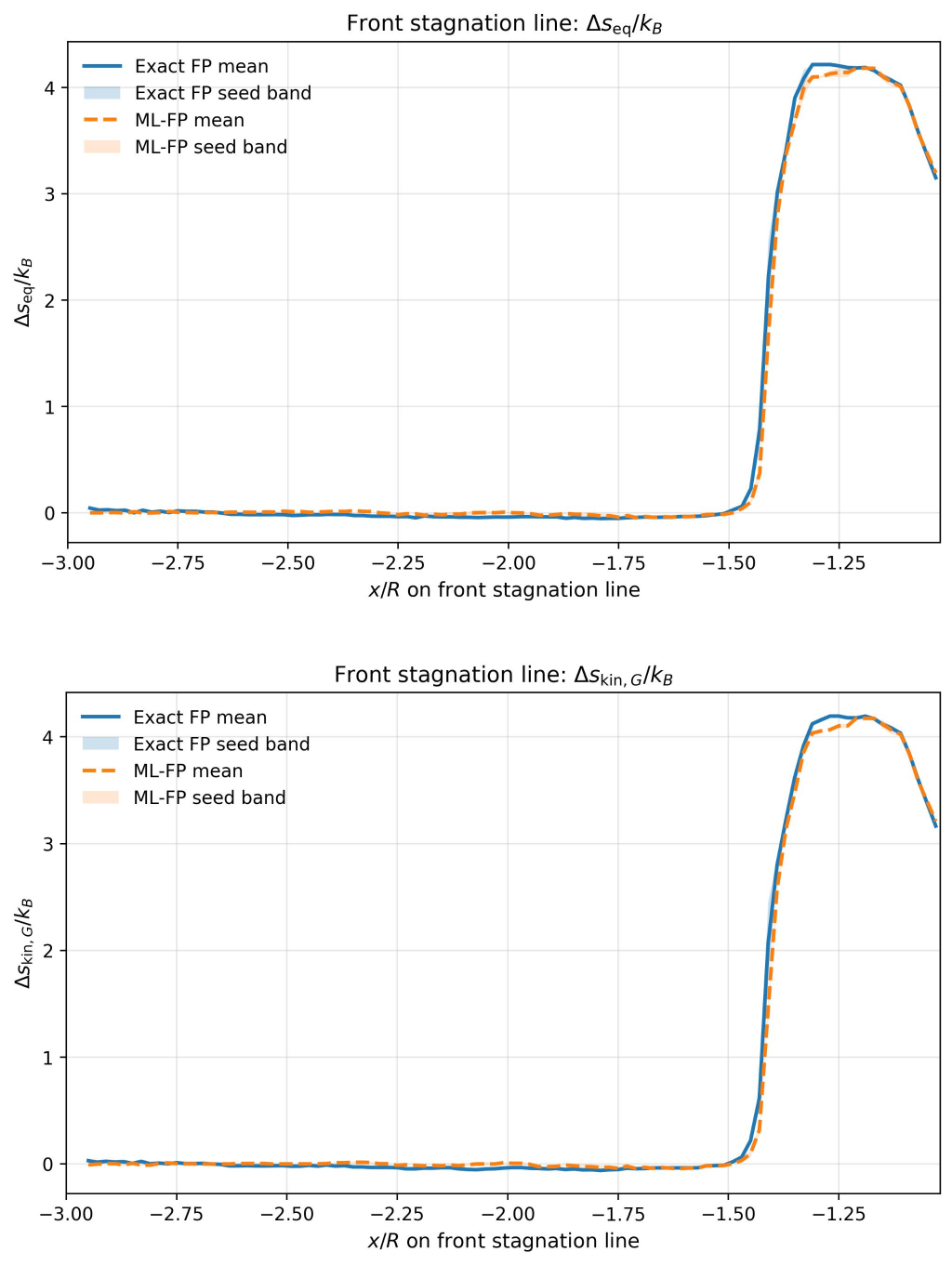}
\caption{\reventropy{Covariance-based entropy-proxy comparison on the front stagnation line for the hypersonic cylinder. The top panel shows the equilibrium entropy proxy $\Delta s_{\rm eq}/k_B$ and the bottom panel shows the Gaussian kinetic entropy proxy $\Delta s_{{\rm kin},G}/k_B$. Solid curves denote the exact cubic-FP two-replica mean, dashed curves denote the ML-FP two-replica mean, and the shaded bands show the min--max range across replicas. These profiles provide the high-support entropy audit through the detached bow shock and the stagnation layer.}}
\label{fig:cylinder_entropy_stag}
\end{figure}

\reventropy{The complete entropy-proxy extraction workflow is summarized in Algorithm~\ref{alg:cylinder_entropy_audit}. The compact profile-error metrics in Table~\ref{tab:cylinder_entropy_metrics} are reported for the front stagnation line shown in Fig.~\ref{fig:cylinder_entropy_stag} and the valid fine-resolution front-side near-wall surface gas layer shown in Fig.~\ref{fig:cylinder_entropy_surface}, where the sampling support is high. The rear/wake coarse-bin extension in Fig.~\ref{fig:cylinder_entropy_surface} is retained as an uncertainty-aware diagnostic display and is not folded into the compact surface-error table.}

\begin{algorithm}[H]
\small
\caption{Cylinder covariance-based entropy-proxy audit used for the exact-FP/ML-FP comparison.}
\label{alg:cylinder_entropy_audit}
\begin{algorithmic}[1]
\For{each exact-FP and ML-FP replica}
  \State Run the cylinder solver to the post-transient sampling window with identical grid, particle count, and sampling schedule.
  \State In each sample, bin active particles on the front stagnation line and in the near-wall gas layer on the cylinder surface.
  \State Compute $\Theta$, $\theta$, $D_G$, and the diagnostic $D_q$ directly from peculiar particle velocities; compute $\rho$ from the solver density field.
  \State Normalize by upstream reference values and compute Eqs.~\eqref{eq:cyl_entropy_eq}--\eqref{eq:cyl_entropy_kinG} for each sample.
  \State For surface displays, retain fine front-side bins for $60^\circ\le\theta\le180^\circ$ and aggregate rear/wake bins for $0^\circ\le\theta<60^\circ$ into support-controlled coarse angular bins with within-bin half-range error bars.
\EndFor
\State Time-average the valid-bin profiles, form exact-FP and ML-FP replica means and seed bands for the fine profiles, and report relative $L_2$, mean absolute, and 95th-percentile absolute errors for the high-support audit regions.
\end{algorithmic}
\end{algorithm}

\begin{table}[H]
\centering
\small
\caption{\reventropy{Covariance-based entropy-proxy errors between the ML-FP surrogate and the exact cubic-FP reference. Errors are computed from the two-replica mean profiles over bins satisfying the valid-sample threshold; they should be interpreted as surrogate-fidelity errors, not entropy-production rates.}}
\label{tab:cylinder_entropy_metrics}
\setlength{\tabcolsep}{4pt}
\begin{tabular}{llccc}
\toprule
Profile & Quantity & Rel. $L_2$ error (\%) & Mean abs. & 95th perc. abs. \\
\midrule
Front stagnation & $\Delta s_{\mathrm{eq}}/k_B$ & 4.78 & 0.04029 & 0.1190 \\
Front stagnation & $D_G$ & 7.75 & 0.001562 & 0.007076 \\
Front stagnation & $\Delta s_{\mathrm{kin},G}/k_B$ & 4.83 & 0.03952 & 0.1226 \\
Cylinder surface & $\Delta s_{\mathrm{eq}}/k_B$ & 1.34 & 0.02913 & 0.06495 \\
Cylinder surface & $D_G$ & 4.82 & 0.001978 & 0.007754 \\
Cylinder surface & $\Delta s_{\mathrm{kin},G}/k_B$ & 1.28 & 0.02792 & 0.06174 \\
\bottomrule
\end{tabular}

\end{table}

\reventropy{The main entropy quantities show close agreement as fidelity metrics. The relative errors are $4.78\%$ and $4.83\%$ for the front-line equilibrium and Gaussian kinetic entropy profiles, respectively, and $1.34\%$ and $1.28\%$ on the valid fine-resolution front-side cylinder-surface sector. The $D_G$ comparison is also close in absolute terms: the mean absolute errors are $1.56\times10^{-3}$ on the front line and $1.98\times10^{-3}$ on the surface. These results do not constitute a formal H-theorem for the cubic-FP model or for the neural surrogate; rather, they show that the deployed surrogate does not introduce a visible entropy-proxy discrepancy relative to the exact cubic-FP reference under the validated particle-resolution regime. The front-line relative error is expected to be sensitive to small bow-shock standoff shifts because a minor displacement across a steep shock profile produces a large pointwise difference; the very small absolute $D_G$ differences are consistent with finite-bin sampling noise in the covariance log-determinant estimator. Agreement of entropy levels is a necessary but not sufficient condition for a local production-rate statement, and we therefore keep the claim limited to exact-FP/ML-FP entropy-proxy fidelity.}

\subsection{Detailed Computational Profiling and Efficiency Analysis}

To quantify the computational mechanism for the hypersonic cylinder case, we profiled the exact cubic-FP and GPU-native ML modes over the same 5000-step configuration. The total wall times were 78.7 s for the exact cubic-FP mode and 20.4 s for the ML mode, corresponding to 15.74 ms/step and 4.08 ms/step, respectively. Table~\ref{tab:kernel_breakdown} reports an accounting whose rows sum to the measured totals. The broad particle-motion/sorting category includes different bookkeeping work in the two modes: the exact closure performs additional gather/sort passes needed for high-order moment assembly, whereas the ML mode uses only the low-order state needed by the network. The AMR entry is shown as an amortized per-step cost over the adaptation interval; it is not the raw cost of a single adaptation call.

\begin{table}[htbp]
    \centering
    \caption{Component-level cylinder runtime profile. AMR is reported as an amortized per-step cost over the adaptation interval. All entries are representative synchronized measurements and sum to the measured total time per step up to rounding.}
    \label{tab:kernel_breakdown}
    \renewcommand{\arraystretch}{1.2}
    \resizebox{\textwidth}{!}{%
    \begin{tabular}{lccc}
        \hline
        \textbf{Computational stage} & \textbf{Exact cubic-FP} & \textbf{ML-FP} & \textbf{Role in speedup} \\
        \hline
        Particle motion, boundary treatment, and shared sorting & 1.05 ms & 1.05 ms & Common particle work \\
        Low-order moment gathering & 1.55 ms & 1.55 ms & Required by both modes \\
        High-order moment gathering and extra closure bookkeeping & 9.12 ms & -- & Removed by ML inputs \\
        Dense local $9\times9$ direct solve & 3.70 ms & -- & Removed by ML inference \\
        GPU-native DNN forward pass & -- & 1.16 ms & Replacement closure evaluation \\
        AMR and grid bookkeeping, amortized & 0.32 ms & 0.32 ms & Not optimized \\
        \hline
        \textbf{Total time/step} & \textbf{15.74 ms} & \textbf{4.08 ms} & \textbf{3.85$\times$ speedup} \\
        \hline
    \end{tabular}%
    }
\end{table}

The main acceleration comes from removing the high-order particle-moment gathering and the dense local direct solve needed by the deterministic closure. The $9\times9$ system itself is small, but assembling its right-hand side and matrix requires fourth- and fifth-order particle statistics in every active cell. The ML mode replaces that entire closure assembly by low-order moments and a GPU-native forward pass. The reported speedup is therefore configuration-dependent: if particle motion, boundary treatment, sorting, or AMR dominate a different configuration, the global speedup will be smaller.

\subsubsection{Configuration-dependent Amdahl bound}
The measured speedup should not be interpreted as a universal theoretical maximum. It is the observed reduction for the present grid, particle count, and adaptation frequency. Once the high-order closure work is removed, the remaining particle-motion, boundary, low-order moment, and grid-management stages set the residual cost. Further acceleration of the DNN alone would therefore have diminishing impact unless the remaining particle and AMR components are also optimized.

\paragraph{Cylinder break-even}
The cylinder online saving is 58.3 s per run for the profiled 5000-step case. As an illustrative estimate, if the same offline training cost used for the cavity surrogate is used for the cylinder deployment, the training-only break-even is approximately 20 cylinder runs and the full data-generation plus training break-even is approximately 55 cylinder runs. In practice, the break-even depends on whether the cylinder training data are generated specifically for one condition or reused across a parameter sweep.

\subsection{Heuristic coefficient-sensitivity probe}
As a heuristic sensitivity probe for the cylinder surrogate, we repeated the GPU-native forward pass $N=50$ times with an artificial dropout mask of rate $p=0.2$ activated during inference. The production deployment itself is deterministic and uses no dropout. Because the network architecture used for the reported production model was not trained with dropout layers, the resulting fields should not be interpreted as calibrated Bayesian epistemic uncertainty in the strict Gal--Ghahramani sense. They are reported only as a coefficient-sensitivity indicator for locating regions where the learned map is most fragile. Let $\mathbf{y}^{(k)}$ denote the output vector of the $k$-th perturbed forward pass. The sensitivity for each coefficient is quantified by the standard deviation across these passes:

\begin{equation}
    \sigma_i = \sqrt{\frac{1}{N} \sum_{k=1}^{N} \left( y_i^{(k)} - \bar{y}_i \right)^2},
\end{equation}

\noindent where $\bar{y}_i$ is the mean prediction. Since the closure involves multiple coupled coefficients for the stress tensor ($\mathbf{C}$) and heat flux vector ($\mathbf{\Gamma}$), we define the aggregate sensitivity fields using the Euclidean norm ($L_2$) of their respective standard deviations:

\begin{equation}
    \sigma_{\text{Stress}} = \sqrt{\sum_{j \in \mathcal{I}_C} \sigma_j^2}, \quad 
    \sigma_{\text{Heat}} = \sqrt{\sum_{j \in \mathcal{I}_\Gamma} \sigma_j^2}.
\end{equation}

\noindent Here, $\mathcal{I}_C$ and $\mathcal{I}_\Gamma$ represent the sets of indices corresponding to the stress and heat flux coefficients in the network output layer, respectively.

\begin{figure}[htbp]
    \centering
    \begin{subfigure}[b]{0.48\textwidth}
        \centering
        \includegraphics[width=\textwidth]{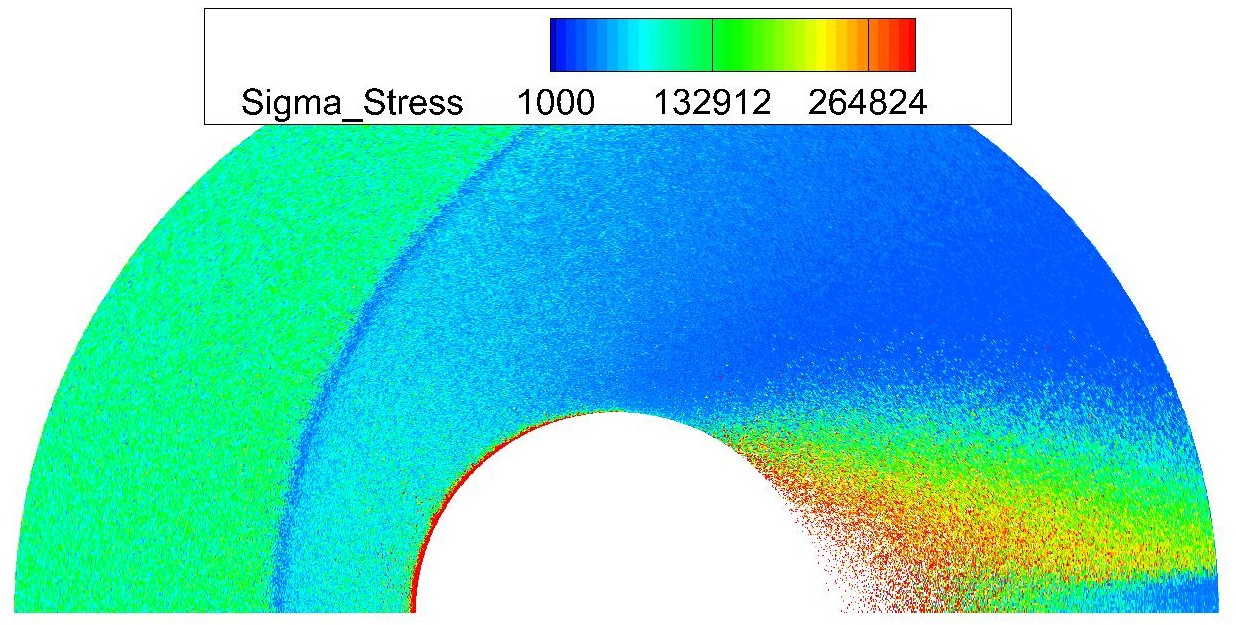}
        \caption{Stress-coefficient sensitivity ($\sigma_{\text{Stress}}$)}
        \label{fig:uq_stress}
    \end{subfigure}
    \hfill
    \begin{subfigure}[b]{0.48\textwidth}
        \centering
        \includegraphics[width=\textwidth]{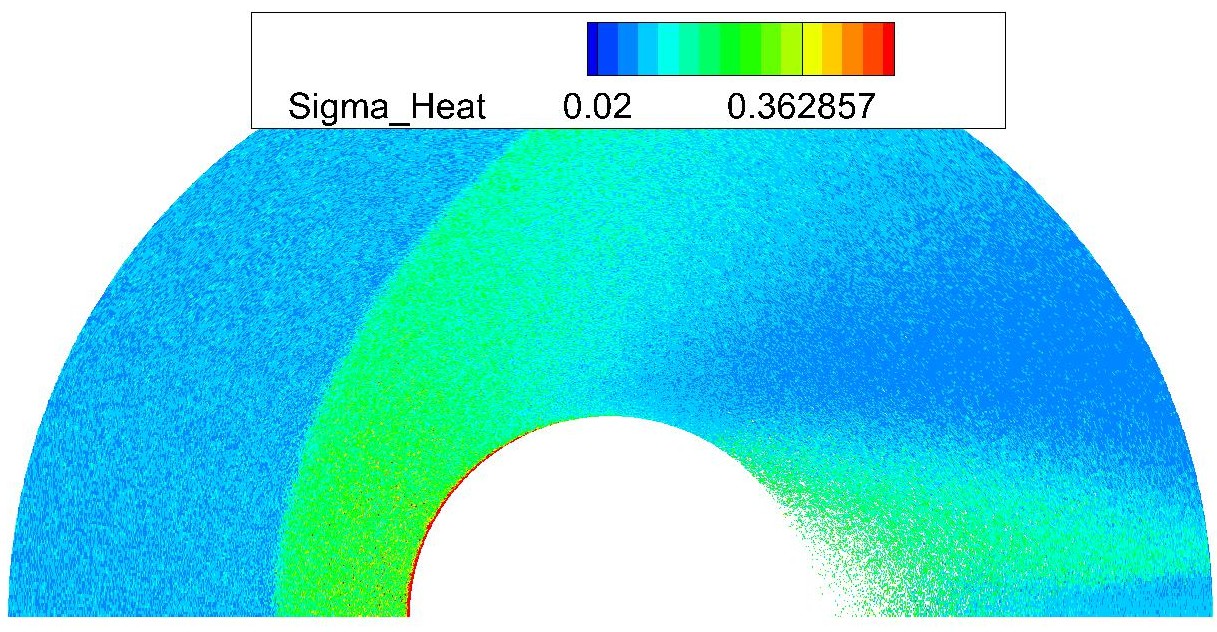}
        \caption{Heat-flux coefficient sensitivity ($\sigma_{\text{Heat}}$)}
        \label{fig:uq_heat}
    \end{subfigure}
    
    \caption{Heuristic coefficient-sensitivity map for the GPU-native surrogate in the cylinder case using inference-time dropout perturbations. The contours show sensitivity of the stress-relaxation and heat-flux coefficient blocks to output perturbations; the highlighted regions coincide with the detached bow shock and stagnation layer. The map is not interpreted as calibrated Bayesian epistemic uncertainty.}
    \label{fig:uq_contours}
\end{figure}

Figure~\ref{fig:uq_contours} illustrates the spatial distribution of this heuristic coefficient sensitivity. The sensitivity is not randomly distributed; rather, it is structurally correlated with the flow physics. The largest sensitivity values (indicated by red and yellow contours) are concentrated in two specific regions:

The Bow shock: The distinct arc-shaped region of high sensitivity follows the location of the detached bow shock. This is expected, as the shock wave represents a strong discontinuity where the velocity distribution function deviates significantly from equilibrium, making the mapping from moments to closure coefficients highly non-linear and challenging for the network.

The stagnation region: High sensitivity is also observed near the stagnation point on the cylinder surface, characterized by extreme compression and thermal gradients.

Conversely, in the free-stream region and the far wake, the sensitivity is minimal (blue), indicating that the deterministic network output is less affected by coefficient perturbations in near-equilibrium regimes. This heuristic map is useful for identifying regions where hybrid physics-ML strategies or mesh refinement might be necessary, but it is not a calibrated Bayesian uncertainty estimate.

\section{Conclusion}

\rev{This work demonstrates a GPU-native neural surrogate for the deterministic closure step in particle-based cubic-Fokker--Planck simulations. The main contribution is not only the neural approximation itself, but its integration directly inside the GPU particle loop without CPU--GPU transfers. The validation emphasizes quantitative evidence and limitations.}

\rev{The Couette problem serves as a compact verification and timing case. The 2D lid-driven cavity is the primary internal-flow benchmark. For the q-weighted cavity surrogate, the ML run closely reproduces the macroscopic fields and preserves the spatial structure of high-order nonequilibrium diagnostics. Additional tests at nominal $Kn=0.5$ and $Kn=1.0$ show that the surrogate remains accurate in moderately rarefied transitional conditions represented by the validated training manifold.}

\rev{The analysis also clarifies the performance mechanism. Component profiling shows that the major removed cost is not only the small $9\times9$ linear solve, but the full high-order moment gathering required to assemble the deterministic closure. The ML mode replaces this by low-order moment gathering and a lightweight GPU-native forward pass. The attainable speedup is therefore configuration-dependent and bounded by the remaining particle motion, boundary treatment, and low-order moment operations. The break-even analysis shows that the offline cost is recovered only after a sufficient number of target simulations, making the method most attractive for repeated-query studies, parametric sweeps, and uncertainty analyses.}

\rev{The paper also addresses physical consistency concerns. The present surrogate does not impose a rigorous H-theorem or a universal realizability projection. Instead, the study reports direct stability, positivity, coefficient-boundedness, a time-averaged closure-coefficient audit, entropy-proxy fidelity, and particle-count sensitivity diagnostics. For the hypersonic cylinder, the covariance-based entropy-fidelity audit computes $\Delta s_{\rm eq}/k_B$, $D_G$, and $\Delta s_{{\rm kin},G}/k_B$ from particle velocity moments and solver density fields in exact-FP and ML-FP replicas. This audit shows that the deployed neural $C/\Gamma$ closure reproduces the front-stagnation and near-wall Gaussian entropy-proxy profiles of the exact cubic-FP solver under the validated particle-resolution regime. It is not an entropy-production or H-theorem proof. No coefficient clipping or projection was required in the reported q-weighted cavity runs. The learned map should be interpreted as valid on the distribution manifold represented by the training simulations; it is not claimed to be universally identifiable from low-order moments in the full kinetic space.}

\rev{For the hypersonic cylinder problem, the manuscript retains the most reliable validation results: bow-shock structure and surface coefficients. Time-averaged high-order cylinder diagnostics were explored, but the corresponding two-dimensional maps are highly sensitive to sparse wake statistics and adaptive-mesh sampling. They are therefore not used as main-text evidence. This conservative choice avoids overstating diagnostics that are dominated by sparse wake sampling and adaptive-grid binning artifacts.}

\rev{Overall, the GPU-native surrogate provides a practical acceleration path for particle-based cubic-FP simulations when the target flow is represented by the training data and when many related simulations are required. Future work will focus on physics-constrained coefficient projections, distribution-aware features that improve identifiability, adaptive retraining for far-from-equilibrium shocks, and a more complete treatment of free-molecular limits where the cubic-FP reference model is no longer adequate.}

\end{document}